\documentclass[12pt,dvipsnames]{article}
\usepackage[top=80pt,bottom=85pt,left=75pt,right=75pt]{geometry}

\pdfoutput=1
\usepackage[utf8]{inputenc}
\usepackage[T1]{fontenc} 
\usepackage[english]{babel} 
\usepackage{cancel}
\usepackage{braket}
\usepackage{csquotes}
\usepackage[dvipsnames]{xcolor} %Package for fancy coloros
\usepackage{physics}
\usepackage{tabularx} 
\usepackage{booktabs} 
\usepackage{dsfont}
\usepackage{rotating}
\usepackage{multirow} 
\usepackage{comment} 
\usepackage{amsmath}
\usepackage{bbm}
\usepackage{tikz}
\usetikzlibrary{math} 
\usetikzlibrary{3d} 
\usepackage{tikz-3dplot}
\usepackage{tikz-cd} 
\usepackage{microtype}
\usepackage{amsthm}
\usepackage{dsfont}
\usepackage[strict]{changepage}

\usepackage{tikz}
\usepackage{asymptote}
\usetikzlibrary{3d}
\usetikzlibrary{calc}
\usepackage{pgfplots}
\usepackage{tikz-3dplot}

\pgfplotsset{
	compat=newest
}

\tikzcdset{
  cells={font=\everymath\expandafter{\the\everymath\displaystyle}},
}

\usepackage{amssymb} 
\usepackage{subcaption} 
\DeclareCaptionFormat{custom}

\usepackage[pdftex,colorlinks=true]{hyperref} 
\hypersetup{urlcolor=MidnightBlue, citecolor=red, linkcolor=MidnightBlue}

\usepackage{float} 
\usepackage[capitalise]{cleveref}

\newcolumntype{E}{>{\hfil$}p{0.65cm}<{$\hfil}}

\newcolumntype{D}{>{\hfil$}p{7.4cm}<{$\hfil}}
\newcolumntype{C}{>{\hfil$}p{3cm}<{$\hfil}}
\newcolumntype{P}{>{\hfil$}p{7.7cm}<{$\hfil}}
\newcolumntype{F}{>{\hfil$}p{5.7cm}<{$\hfil}}
\newcolumntype{L}{>{\hfil$}p{2.6cm}<{$\hfil}}
\newcolumntype{S}{>{\hfil$}p{1.8cm}<{$\hfil}}
\newcolumntype{R}{>{\hfil$}p{5.2cm}<{$\hfil}}
\newcolumntype{U}{>{\hfil$}p{4.2cm}<{$\hfil}}
\newcolumntype{Q}{>{\hfil$}p{6.4cm}<{$\hfil}}
\newcolumntype{T}{>{\hfil$}p{1.9cm}<{$\hfil}}
\newcolumntype{V}{>{\hfil$}p{5.8cm}<{$\hfil}}
\newcolumntype{H}{>{\hfil$}p{1.8cm}<{$\hfil}}
\newcolumntype{A}{>{\hfil$}p{6cm}<{$\hfil}}
\newcolumntype{B}{>{\hfil$}p{2cm}<{$\hfil}}
\makeatletter
\newcommand\xleftrightarrow[2][]{%
  \ext@arrow 9999{\longleftrightarrowfill@}{#1}{#2}}
\newcommand\longleftrightarrowfill@{%
  \arrowfill@\leftarrow\relbar\rightarrow}
\makeatother

\renewcommand{\vec}[1]{\mathbf{#1}}

\definecolor{bittersweet}{rgb}{1.0, 0.44, 0.37}

\numberwithin{equation}{section}

\usepackage[backend=biber,sorting=none,maxnames=99,giveninits=true,doi=false,citestyle=numeric-comp]{biblatex}
\allowdisplaybreaks[1] %Display break of the formulas

\hypersetup{
pdftitle={2312.13360},    % title
pdfauthor={\textcopyright\ Guillermo Arias Tamargo, Mario de Marco},     % author
pdfsubject={HEP},   % subject of the document
pdfcreator={pdfLaTex},   % creator of the document
pdfproducer={LaTex}, % producer of the document
pdfkeywords={},
colorlinks=true,
linkcolor=blue,
urlcolor=blue,
filecolor=blue,
citecolor=blue,
linktocpage=true
}

\definecolor{cambridgeblue}{rgb}{0.64, 0.76, 0.68}
\definecolor{caribbeangreen}{rgb}{0.0, 0.8, 0.6}
\definecolor{celadon}{rgb}{0.67, 0.88, 0.69}
\definecolor{champagne}{rgb}{0.97, 0.91, 0.81}
\definecolor{cream}{rgb}{1.0, 0.99, 0.82}
\definecolor{cyan(process)}{rgb}{0.0, 0.72, 0.92}
\definecolor{brilliantlavender}{rgb}{0.96, 0.73, 1.0}
\definecolor{candypink}{rgb}{0.89, 0.44, 0.48}

\begin{document}

\begin{titlepage}

\begin{flushright}\footnotesize

\texttt{}
\vspace{0.6cm}
\end{flushright}

\mbox{}
\vspace{0truecm}
\linespread{1.1}

%%%%%%%%%%%%%%%%%
\centerline{\LARGE \bf Disconnected gauge groups in the infrared}
\medskip

\vspace{1.5truecm}

\centerline{
    { Guillermo Arias-Tamargo}\textsuperscript{\it a} \footnote{guillermo.arias.tam@gmail.com}, and
    { Mario De Marco}\textsuperscript{\it b} \footnote{mario.demarco@math.uu.se}}

\vspace{1cm}
\centerline{\textsuperscript{\it a}\it  Theoretical Physics Group, The Blackett Laboratory, Imperial College London,}

\centerline{\it  Prince Consort Road London, SW7 2AZ, UK}
\medskip
\centerline{{\textsuperscript{\it b}\it  Mathematics Institute, Uppsala University, \\ Box 480, SE-75106 Uppsala, Sweden\\}}
\medskip
\vspace{1cm}

\centerline{\small{\bf Abstract} }

\begin{center}
\begin{minipage}[h]{0.85\textwidth}
\small
Gauging a discrete 0-form symmetry of a QFT is a procedure that changes the global form of the gauge group but not its perturbative dynamics. In this work, we study the Seiberg-Witten solution of theories resulting from the gauging of charge conjugation in 4d $\mathcal N = 2$ theories with $SU(N)$ gauge group and fundamental hypermultiplets. The basic idea of our procedure is to identify the $\mathbb{Z}_2$ action at the level of the SW curve and perform the quotient, and it should also be applicable to non-lagrangian theories. We study dynamical aspects of these theories such as their moduli space singularities and the corresponding physics; in particular, we explore the complex structure singularity arising from the quotient procedure. We also discuss some implications of our work in regards to three problems: the geometric classification of 4d SCFTs, the study of non-invertible symmetries from the SW geometry, and the String Theory engineering of theories with disconnected gauge groups.
 
\end{minipage}

\end{center}
\end{titlepage}
\newpage

\tableofcontents 

\setcounter{footnote}{0} 
\setcounter{page}{1}
\section{Introduction and summary}

In recent years, it has been appreciated that when studying the behavior of a Quantum Field Theory (QFT), the global form of the gauge group, and not only its Lie algebra, can play a very important role \cite{Aharony:2013hda}. A classic example is the case of $SU(2)$ and $SO(3)$, which have a different spectrum of line operators, periodicity of the $\theta$ angle, etc.; which in turn translate into stark differences in their IR dynamics. In the modern language of generalised global symmetries \cite{Gaiotto:2014kfa} it is their differences in higher form symmetries and anomalies that gives rise to the change in behaviour. In this particular example, the different ``global form'' of the gauge group refers to the fundamental group $\pi_1(G)$ and the two QFTs are related by gauging a (discrete) 1-form symmetry.

In this work, we shall be focusing on a slightly different situation, namely, we will concern ourselves with the gauging of a discrete 0-form symmetry, leading to a different $\pi_0(G)$ for the gauge group, that displays now disconnected components. Examples of this type of operation feature frequently in theoretical physics, from the GSO projection in String Theory to various dualities in lattice systems (see e.g. \cite{Tachikawa:2017gyf,Karch:2019lnn,Seiberg:2023cdc}). Quantum Gravity considerations also lead in a natural way to disconnected gauge groups: we know that our world is described at low energies by some gauge theory, and very often these have discrete global symmetries. But we also know that global symmetries in quantum gravity, including discrete ones, should be either broken or gauged when going to high energies \cite{Kallosh:1995hi,Banks_2011,Harlow:2018tng,Heidenreich:2021xpr,Harlow:2023hjb}, which in the latter case will lead us to said type of group. In particular, we will study the case where the discrete symmetry we gauge corresponds to charge conjugation.

Charge conjugation has the particularity that its action doesn't commute with a gauge transformation. As a global symmetry, this has the implication that the total symmetry of the theory under consideration will form a non-trivial 2-group structure \cite{Benini:2018reh,Bhardwaj:2022scy}. When gauging it, it means that the total gauge group will be a semidirect product. This has several interesting consequences. In \cite{Arias-Tamargo:2022nlf,Bhardwaj:2022yxj,Heidenreich:2021xpr} (and similarly in spirit \cite{Antinucci:2022eat,Damia:2023gtc}) it was shown that some of the topological defects of the theory are acted upon by charge conjugation, and thus the gauge invariant combination one can build from them is non-invertible in any dimension (see \cite{Shao:2023gho,Schafer-Nameki:2023jdn,Bhardwaj:2023kri,Carqueville:2023jhb,Brennan:2023mmt} for recent reviews of this fascinating and fast-developing subject). The spectrum of gauge invariant local operators is also modified; as an example, the baryon and anti-baryon of $SU(N)+N_f$ fundamentals become the same local operator after gauging charge conjugation. When restricting to the case of supersymmetric gauge theories with 8 supercharges, tools such as indices or Hilbert Series allow for an in-depth study of the spectrum of local operators at the quantum level \cite{Bourget:2018ond,Argyres:2018wxu,Arias-Tamargo:2019jyh,Arias-Tamargo:2021ppf}. Indeed, these works showed that the disconnected gauge groups provide the first known examples of 4d $\mathcal{N}=2$ SCFTs whose Coulomb branch is not freely generated.

The discussion above concerns only kinematical aspects of QFTs with disconnected gauge groups. Our aim in this work is to begin the investigation of their dynamics. A zero-th order observation in this regard is that since the discrete gauging is changing the global form of the gauge group but not its Lie algebra, all perturbative quantum corrections (Feynman diagrams, etc.) should be identical. Instead, something more interesting might come by looking at non-perturbative phenomena. Once again, we will invoke supersymmetry as a simplifying assumption that will give us the tools necessary to address this question in an exact way.

In the seminal work by Seiberg and Witten \cite{Seiberg:1994rs,Seiberg:1994aj} it was found that 4d $\mathcal{N}=2$ SUSY imposes sufficient constraints on the theory so that the low energy EFT in the Coulomb branch can be computed exactly. The idea is that the metric entering the kinetic term of the vector multiplet scalars at low energy is encoded in an auxiliary object, the Seiberg-Witten (SW) curve and differential (see e.g. \cite{Lerche_1997, Tachikawa:2013kta,Akhond:2021xio, Martone:2020hvy} for pedagogical reviews of the topic). In the example of pure $SU(2)$ SYM, the curve and differential would be 
\begin{align}
    y^2 = (x^2 + u_2)^2 - 4 \Lambda^4,\qquad \lambda = x\, d \left[\log\left(\frac{x^2 + u_2-y}{x^2 + u_2 + y}\right)\right]
\end{align}
where $u_2$ is the Coulomb branch parameter, $(x,y)$ are the complex variables parametrizing the curve and differential, and $\Lambda$ is a scale involving the UV cutoff and gauge coupling. From these, one computes the periods $a$ and $a_D$ by integrating the SW differential over the $A$ and $B$ cycles of the curve, and from those one finds the metric of the low energy NLSM (which in the $SU(2)$ would have only one component),
\begin{align}
    \tau (a) = \frac{\partial a_D}{\partial a}\,. 
\end{align}
This metric has singularities whenever the SW curve on top of the Coulomb branch has a degeneration. The interpretation is that the low energy EFT fails because there is an extra particle becoming massless. Famously, in the $SU(2)$ case these singularities occur at $u=\pm 2 \Lambda^2$ and correspond to a monopole and a dyon becoming massless. 

The question we ask is: how are the SW curve and differential modified after the discrete gauging of a 0-form symmetry? This problem was addressed in \cite{Argyres:2016yzz} in the rank 1 case (see also \cite{Furrer:2024zzu,Giacomelli:2024sex} for more recent work). The basic idea is that the extension of the gauge group results in a quotient of the moduli space (morally speaking, ``doubling the size'' of the gauge orbits results in ``half the amount'' of physically distinct configurations). From the lagrangian, we know how the symmetry acts on the Coulomb branch VEVs and how to take the quotient. But the full SW solution is given by a fibration over the Coulomb branch, and therefore to be able to consistently take the quotient we need to find the correct gluing isomorphism that identifies the fibers on top of the corresponding points in the parent theory Coulomb branch $\mathcal M_{\text{CB}}$. In \cite{Argyres:2016yzz} it was observed that for the rank 1 case the only discrete 0-form symmetries one can gauge are combinations of $U(1)_R$ R-symmetry, $SL(2,\mathbb{Z})$ (for fixed values of the coupling) and the of outer automorphism of the flavor group $\text{Out}(F)$. Our starting point will be instead $SU(N)$ SQCD theories for higher ranks, which has an extra symmetry not present at rank 1, the outer automorphism of the gauge group $\text{Out}(G)$, whose gauging is compatible with supersymmetry \cite{Bourget:2018ond, Argyres:2018wxu}. This is a $\mathbb{Z}_2$ that acts as charge conjugation, and that leads us to the disconnected group $SU(N) \rtimes \mathbb{Z}_2$ and which henceforth we will be denoting as $\widetilde{SU}(N)$. 

Let us anticipate what are the main results of our analysis. We are indeed able to compute the SW solution for $\widetilde{SU}(N)$ theory, focusing for simplicity in either the pure SYM case or the conformal case (although from the latter we can straightforwardly integrate out part of the flavours introducing the corresponding mass parameter in the curve and taking the appropriate decoupling limit). We are not able to give a closed expression for generic $N$, because there appear to be several sources of non-generic-ness, as we will discuss in the main text. The most important one is that, from the $SU(4)$ case on, the Seiberg-Witten geometry is not a complete intersection anymore. Nevertheless, our procedure is straightforward to generalise to arbitrary rank. As checks of our proposals, we are able to match the global symmetry seen by the curve with the expectation from the lagrangian, as well as the monodromy at infinity due to the 1-loop $\beta$-function. 

We also identify the BPS particles that become massless at different points of the Coulomb branch by studying the vanishing cycles of our SW curves. Once again, we cannot perform the analysis for generic $N$, but we do discuss the universal features that we expect to hold in all cases. Most surprisingly, we find the existence of a new singularity in the origin of the moduli space, which does not descend from a massless particle of $SU(N)$ before the discrete gauging, and which plays a very important role in making the solution consistent. Moreover, using the fact that our method for analyzing the spectrum is purely geometrical, we will be able to (conjecturally) extend this analysis beyond the $\mathbb{Z}_2$ gauging in the $SU(N) + F$ case. Indeed, under some technical assumptions, we will describe the effect of the discrete-gauging of a $\mathbb Z_k$ zero-form symmetry on the BPS spectrum of a generic, not necessarily lagrangian, 4$d$ $\mathcal N =2$ theory.\footnote{We should remark that for these cases the consistency of the discrete gauging with supersymmetry is an assumption rather than known from a lagrangian construction.}

There are three particular problems where our results can help to shed some light. The first concerns the bottom-up classification of 4d $\mathcal{N}=2$ SCFTs. In the rank 1 case \cite{Argyres:2015ffa,Argyres:2015gha,Argyres:2016xua,Argyres:2016xmc} it turned out that several of the allowed SW geometries correspond to discrete gaugings \cite{Argyres:2016yzz}. At rank 2 \cite{Martone:2020nsy, Argyres:2020wmq,Argyres:2022lah,Argyres:2022puv,Argyres:2022fwy} (see also \cite{Caorsi:2018zsq,Xie:2022aad,Li:2022njl,Xie:2023lko,Xie:2023wqx}) a puzzle occurs: the scaling dimensions of the Coulomb branch parameters of $\widetilde{SU}(3)$ with 6 fundamentals are $(2,6)$ \cite{Bourget:2018ond,Argyres:2018wxu,Arias-Tamargo:2019jyh} and the flavor group is of rank three, which at a first glance is not allowed \cite{Argyres:2022lah}. Interestingly, it's the aforementioned new singularity at the origin of the Coulomb branch that explains how the $\widetilde{SU}(3)$ theory relates to the classification of Argyres and Martone.

The second problem concerns the study of generalized global symmetries with the formalism of the Symmetry Topological Field Theory (SymTFT) \cite{Apruzzi:2021nmk,DelZotto:2022joo,Apruzzi:2022dlm,DelZotto:2022ras,Hosseini:2021ged,Argyres:2022kon,Kaidi:2022cpf,vanBeest:2022fss,Antinucci:2022vyk,Kaidi:2023maf,Apruzzi:2023uma,Sacchi:2023omn,Chen:2023qnv,Bhardwaj:2023ayw}. This is an auxiliary topological field theory in one dimension higher generalising the concept of anomaly inflow, that encodes all the symmetries of the theory of interest. In \cite{DelZotto:2022ras} it was shown how the SymTFT for the 1-form symmetry can be obtained from the Seiberg-Witten solution, which has the implication that the deep IR knows about the fundamental group of the gauge group (see also \cite{Closset:2023pmc}). Given that the theories we study have a non-invertible 1-form symmetry, it's interesting to check whether the method of \cite{DelZotto:2022ras} (which amounts to finding the BPS particles that can screen the relevant line operators) can detect it: we find that the answer is no, and that the non-invertibility should be otherwise encoded in the geometry.

The third one concerns the String Theory engineering of theories with disconnected gauge groups. As showed by Witten \cite{Witten:1997sc}, there is an intimate relation between the Seiberg-Witten solution and the geometry of the type IIA or M-theory setup that engineers the theory. Starting from a setup of M5 branes wrapping a curve $\mathcal{C}$ (by specifying the curve one is specifying the theory as in the class $\mathcal{S}$ construction \cite{Gaiotto:2009we}), in the IR the system will flow to a single M5 wrapping a certain cover of $\mathcal{C}$ which is precisely the Seiberg-Witten curve. In principle, it is easy to reverse engineer this process in order to find what is the UV curve wrapped by the M5 branes from the SW solution; however, as we shall see, there is an additional complication for the disconnected gauge groups.

The rest of this paper is organized as follows. In section \ref{sec:lagrangian} we review the lagrangian construction of $\mathcal{N}=2$ SYM theories based on disconnected gauge groups. In section \ref{sec:curve} we obtain their Seiberg-Witten geometry, including some checks for the validity of our proposal. In section \ref{sec:bps_quiver} we compute the spectrum of particles that become massless at the various singularities in the Coulomb branch, and identify their electric and magnetic charges from the monodromies in the SW curve. We finish in section \ref{sec:conclusion} with a discussion of our results in regards to the three aforementioned problems: the classification of rank 2 geometries, the derivation of the SymTFT from the IR, and the uplift of the SW geometry to M-theory; as well as some additional open questions. We relegate to the appendices some auxiliary computations. In appendix \ref{sec:SW_su2} we reproduce the already known results in the rank 1 case of $SU(2)+4 F$ using our (slightly different) methods. In appendix \ref{sec:app_monodromies} we give the details of the proof of consistency of the monodromies in our curves.

\section{Discrete gauging: UV perspective}\label{sec:lagrangian}

The main goal of this work is to explore the consequences of the discrete gauging of the outer automorphism of the gauge group in the Seiberg-Witten geometry of the 4d $\mathcal{N}=2$ SYM theory with $SU(N)$ gauge group and fundamental flavours. With this in mind, the first step is to understand the UV lagrangian. Our point of view will be to consider the disconnected Lie group built as an extension of $SU(N)$ by its outer automorphism, which we denote $\widetilde{SU}(N)$, and then study the supersymmetric gauge theories we can build with it. This is equivalent to starting with a $SU(N)$ (supersymmetric) gauge theory and gauging the 0-form global symmetry corresponding to its outer automorphism.

\subsection{Gauging the outer automorphism of \texorpdfstring{$SU(N)$}{}}

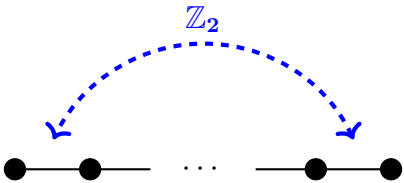
\begin{figure}[t]
    \centering
\begin{tikzpicture}
    \draw[black,fill] (0,0) circle (4pt);
    \draw[black,fill] (1,0) circle (4pt);
    \draw[black,fill] (4,0) circle (4pt);
    \draw[black,fill] (5,0) circle (4pt);
    \node at (2.5,0) {$\cdots$};
    \draw[black, thick] (0,0)--(1.8,0);
    \draw[black, thick] (3.2,0)--(5,0);  
    \draw[blue,dashed,ultra thick, <->] (4.5,0.4) arc (25:155:2.2);
    \node at (2.5,2) {\textcolor{blue}{$\bf \mathbb{Z}_2$}};
\end{tikzpicture}
    \caption{The outer automorphism of the gauge group $SU(N)$ is a (0-form) global symmetry of the gauge theory when the rank is higher than 1. Intuitively it corresponds to charge conjugation, as matter in the fundamental representation is mapped to the antifundamental (and viceversa).}
    \label{fig:outer_suN}
\end{figure}

There are in principle infinitely many ways to build a group extension of $SU(N)$ by any other group $G$, specified by building a short exact sequence of the form
\begin{align}
    1 \to SU(N) \to \widetilde{G} \to G \to 1 \,.
\end{align}
We will focus on the very simple case where the group we use to build the extension is $\mathbb{Z}_2$ and the sequence splits,
\begin{align}
    1 \to SU(N) \rightleftarrows \widetilde{SU}(N) \rightleftarrows \mathbb{Z}_2 \to 1 \,.
\end{align}
In order words, $\widetilde{SU}(N)$ is a semidirect product,
\begin{align}\label{eq:sutilde_semidirect}
    \widetilde{SU}(N) = SU(N) \rtimes_\Theta \mathbb{Z}_2\,.
\end{align}
Moreover, we can identify the $\mathbb{Z}_2$ with the group of outer automorphisms of $SU(N)$. This is the symmetry of the Dynkin diagram of $SU(N)$ that acts by exchanging the roots $\alpha_j \leftrightarrow \alpha_{N-1-j}$ (see Figure \ref{fig:outer_suN}). This action on the root system can be lifted to the group and used to specify the homomorphism $\Theta$ involved in the definition of the semidirect product \eqref{eq:sutilde_semidirect},\footnote{A detailed analysis of all the possible ways to extend $SU(N)$ by its group of outer automorphisms was carried out in \cite{Arias-Tamargo:2019jyh}. It was found that if $N$ is odd, $\Theta$ is uniquely defined; while if $N$ is even, there are two possibilities, one of which coincides with the $N$ odd case. The two extensions were dubbed $\widetilde{SU}(N)_{I,II}$. When analysing the moduli space of 4d $\mathcal{N}=2$ theories with these disconnected gauge groups, it was found that while the Higgs branch does depend on the choice of extension, the Coulomb branch does not (see also \cite{Arias-Tamargo:2021ppf,Bourget:2018ond}). Therefore we will restrict to the type $I$ extension common to both $N$ even and odd, which is defined by the homomorphism \eqref{eq:homomorphism_semidirect}. }
\begin{align}\label{eq:homomorphism_semidirect}
    \Theta_1(g) = g\,,\quad \Theta_{-1}(g) = (g^{-1})^T=g^*\,,
\end{align}
where $g\in SU(N)$. Elements in $\widetilde{SU}(N)$ can be described as pairs $(g,\epsilon)$ with $g\in SU(N)$ and $\epsilon\in\mathbb{Z}_2$, with the product given by
\begin{align}
    (g_1,\epsilon_1)\cdot (g_2,\epsilon_2) = (g_1 \Theta_{\epsilon_1}(g_2),\epsilon_1 \epsilon_2)\,.
\end{align}

Sometimes it is useful to use directly the fundamental representation, which has dimension $2N$ and where the two connected components of the group can be written as
\begin{align}\label{eq:matrix_repr}
    \widetilde{SU}(N) = \left\lbrace \left(\begin{array}{cc}
        g & 0 \\
        0 & g^*
    \end{array}\right),\, g\in SU(N) \right\rbrace \cup \left\lbrace \left(\begin{array}{cc}
        0 & g \\
        g^* & 0
    \end{array}\right) ,\, g \in SU(N) \right\rbrace\,,
\end{align}
and where the element that generates the $\mathbb{Z}_2$ and which takes us from one connected component of the group to the other is
\begin{align}\label{eq:generator_z2_fund}
    (\text{Id},-1) \to \left(\begin{array}{cc}
        0 & \mathds{1}_N \\
        \mathds{1}_N & 0
    \end{array}\right)\,.
\end{align}

We will also need to use the adjoint representation. In this case, as opposed to the fundamental, we don't need to double the size of the matrices when going to the extended group. This can be seen to be the case because the Dynkin labels of the adjoint of $SU(N)$ are $[1,0,\dots,0,1]$ which is symmetric under the $\mathbb{Z}_2$ outer automorphism (see \cite[section VI]{brocker2013representations}). Instead, the adjoint of the connected group survives directly in $\widetilde{SU}(N)$. The consequence of the extension is that we have another $N^2-1$ dimensional representation, built by tensoring the adjoint with the fundamental of the $\mathbb{Z}_2$, called the pseudo-adjoint (by analogy with the pseudo-vector representation of $O(N)$ groups); however, we will not need it in the rest of the paper.

Before moving on to building theories based on these gauge groups, there are two important remarks that we'd like to make. The first is that, as is manifest by the matrix representation \eqref{eq:matrix_repr}, one can exchange the roles of $g$ and $g^*$ by multiplication with elements of $\widetilde{SU}(N)$ (concretely, by conjugation with the element \eqref{eq:generator_z2_fund}). In this sense we can understand that when building the group extension we are gauging charge conjugation. The second is that, while the fundamental representation of $SU(N)$ is complex, after the discrete gauging, the new fundamental representation \eqref{eq:matrix_repr} is real; a fact that will become important to build lagrangians in the following section. An intuitive way to understand this is to think of this representation as the one corresponding to the Dynkin labels $[1,0,\dots,0]\oplus [0,\dots,0,1] $, except it's made irreducible by the $\mathbb{Z}_2$ that mixes the two summands. One can also rigorously check the reality of this representation by e.g. computing the Frobenius-Schur indicator \cite{Arias-Tamargo:2019jyh}.

\subsection{Matter fields, lagrangians and global symmetry}

The usual way to construct an $\mathcal{N}=2$ supersymmetric gauge theory is to start with $\mathcal{N}=1$ superfields and lagrangian, and then impose the necessary restrictions so that one has extended supersymmetry (usually by means of looking at the $R$-symmetry). Let's focus first on the hypermultiplets. The story is completely analogous to the textbook case of $SU(2)$: we start with a collection of two different $\mathcal{N}=1$ chiral superfields
\begin{align}
    Q_I^a\,,\quad \widetilde{Q}^I_a\,,
\end{align}
where $a=1,\dots,N$ is a gauge index and $I$ a flavor index. But, since the fundamental representation of $\widetilde{SU}(N)$, as presented above, is real, we can just  reorganize them as
\begin{align}\label{eq:hypers_indices}
    Q_I^a\,,\quad a=1,\dots,2N\,,\text{ and } I=1,\dots N_f\,,
\end{align}
with $N_f$ even. Again due to the reality of the representation, we have an invariant tensor $\delta_{ab}$ which is symmetric. We can use it to build the quadratic term in the superpotential.
\begin{align}\label{eq:W_cuadratic}
    W \, \supset \, M^{IJ}\, Q^a_I \, Q^b_J \, \delta_{ab}\,,
\end{align}
where the mass matrix satisfies $M^{IJ}=M^{JI}$. From this, we already see that SYM with $N_f$ flavours in the fundamental of $\widetilde{SU}(N)$ has a $USp(N_f)$ global symmetry.

In order to build the cubic term of the superpotential, we need also a vector multiplet. It is straightforward to construct the corresponding superfields transforming in the adjoint representation, since as we saw in the previous section it trivially survives the discrete gauging. In order to be able to contract the indices with the presentation \eqref{eq:hypers_indices} we construct the following block-diagonal matrix,
\begin{align}\label{eq:doubling_adjoint}
    (\Phi)_{ab} = \left(\begin{array}{cc}
        \varphi &  0 \\
        0 & \varphi^* 
    \end{array}\right)\,,
\end{align}
where $\varphi$ is the $\mathcal{N}=1$ chiral in the $SU(N)$ vector multiplet and now $\Phi$ is a $2N\times 2N$ matrix. Now we can build the cubic term in the superpotential,
\begin{align}\label{eq:W_cubic}
    W \, \supset \, g^{IJ}\, Q^a_I \, \Phi_{ab}\, Q^b_J \,.
\end{align}
We have introduced a matrix of couplings $g^{IJ}$ whose values are fixed by $\mathcal{N}=2$ supersymmetry. The matrix couplings satisfy $g^{IJ}=-g^{JI}$ as a consequence of $\Phi_{ab}$ being antisymmetric, again indicating a $USp(N_f)$ global symmetry.

Now that we have the superpotential $W$, the only thing left to write down the complete lagrangian are the kinetic terms. The gauge kinetic term is completely identical to the $SU(N)$ case, and the same happens for the Kahler potential of the chiral part of the vector multiplet. Once again, this is because the adjoint representation of $SU(N)$ and $\widetilde{SU}(N)$ are the same; and indeed it was to be expected that anything involving the gauge field, which is a connection taking values on the Lie \emph{algebra} of the group, is not sensitive to the discrete gauging. Lastly, we also write the Kahler potential for the hypermultiplets in the standard fashion, after taking two copies of the $\mathcal{N}=1$ vector superfield analogously to \eqref{eq:doubling_adjoint},
\begin{align}
    K = Q^a_I\,  e^{V_{ab}}\,  Q^b_J \, \delta^{IJ}\,.
\end{align}

To summarize, the only difference between $SU(N)$ and $\widetilde{SU}(N)$ $\mathcal{N}=2$ SYM theories lies in the reality properties of the fundamental representation. From the $F$ and the $\overline{F}$ of $SU(N)$ we build the fundamental of $\widetilde{SU}(N)$, which we'll denote as $(F\oplus\overline{F})$ (keeping in mind that it's irreducible), which requires reorganizing the fields as in \eqref{eq:hypers_indices} and \eqref{eq:doubling_adjoint}. It is important to note that the number of degrees of freedom doesn't change with this reorganization, as it should be with a discrete gauging.

There are two main aspects of the discussion of the lagrangian that will become very important when finding the SW solution. The first is the global symmetry. In $SU(N) + N_f \,F$ SYM, the global symmetry is $U(N_f)$. As we saw, after the discrete gauging, the meson matrix appearing in the quadratic term of the superpotential becomes symmetric, and the global symmetry of $\widetilde{SU}(N)+N_f \,(F\oplus\overline{F})$ is reduced to $USp(N_f)$. Since the SW solution is sensitive to the rank of the global symmetry, we will use this as one of the checks of our geometries. The second concerns the perturbative renormalization of the gauge coupling. As we just mentioned, the gauge field sees only the Lie algebra, and as a result all Feynman rules and loop diagrams are the same after the discrete gauging. This means that the 1-loop correction to the gauge coupling is identical in $SU$ and $\widetilde{SU}$, which will fix for us the monodromy at infinity in the SW curve. It also has the consequence that the condition for the vanishing of the $\beta$-function is not modified, hence $\widetilde{SU}(N)+\, 2N \,(F\oplus\overline{F})$ is superconformal.

\section{Discrete gauging: IR perspective}\label{sec:curve}

Now that we know the lagrangian, we are in a position to study the Seiberg-Witten geometries for the theories with $\widetilde{SU}(N)$ gauge groups. We will begin by giving, in \cref{sec:generalstrategyforswcurves}, the general prescription, that is based on considerations coming just from the rigidity of the IR effective theory (in fact, they could be applied also to the non-Lagrangian case). We will then use this method, in the following subsections, to obtain the curve and differential for many discrete gaugings $SU(N) + F \to \widetilde{SU}(N) + (F\oplus \overline{F})$, concentrating on the cases $N = 3,4,5$. Finally, in \cref{sec:higherranks}, we will give our general predictions on the features of Seiberg-Witten curves for the generic $SU(N) + F \to \widetilde{SU}(N) + (F\oplus \overline{F})$ discrete-gauging.
\subsection{General strategy}
\label{sec:generalstrategyforswcurves}
There are two ingredients that we need in order to compute the SW solution for $\widetilde{SU}(N)$ theories: the solution for the theory before the discrete gauging (following the terminology of \cite{Argyres:2016yzz}, we will sometimes refer to it as the ``parent'' theory), and the action of the $\mathbb{Z}_2$ on the Coulomb branch.\\
\indent In what follows, we will denote with $r$ the rank of the considered theory, with $\mathcal M_{\text{CB}} \cong \mathbb C^r$ the Coulomb branch of the parent theory, and with $\widetilde{\mathcal M_{\text{CB}}}$ the Coulomb branch of the daughter theory.

\begin{itemize}
    \item The SW solution of the $SU(N)$ theory with $N_f=2N$ fundamental flavors is well known \cite{Argyres:1995wt,Hanany:1995na}, 
\begin{equation}
\label{eq:swcurvesunfflavorsconf}
    y^2 = P_N^2(x) - f \prod_{j=1}^{2N}(x + g \mu+ \hat{m_i}), \qquad \lambda = \Big(x + (g-1)\mu\Big) d \left[\log\left(\frac{P_N(x)-y}{P_N(x) + y}\right)\right]\,.
\end{equation}
Here, $(x,y)$ are the complex variables that parametrize the curve,  $\hat{m_i}$ and $\mu \equiv \frac{1}{2N} \sum_{i=1}^{2N} \hat{m_i}$ are the masses of the fundamental flavours, $f = 1-g^2$, $g = g(\tau)$ is a function of the UV coupling $\tau$ (whose explicit form in terms of Jacobi $\theta$-functions we won't need, but we reproduce here for convenience),
\begin{align}
    g(\tau) = \frac{\theta_2(0,\tau)^4 + \theta_1(0,\tau)^4}{\theta_2(0,\tau)^4 - \theta_1(0,\tau)^4}\,,
\end{align}
and $P_N(x)$ is the characteristic polynomial of the VEV of the adjoint scalar in the vector multiplet that parametrizes the Coulomb branch,
\begin{align}
    P_N (x)= \text{det}\left(\mathds{1}x - \langle\Phi\rangle\right)\,.
\end{align}
For convenience, we will often repackage the masses as $m_{i} \equiv g \mu + \hat{m_i}$, with now $\sum_{i} m_i \neq 0$.\\
\indent We will also need the SW data for the $SU(N)$ pure SYM theories \cite{Argyres:1995wt,Hanany:1995na}, 
\begin{equation}
\label{eq:swcurvesunsymgeneral}
    y^2 = P_N^2(x) - 4 \Lambda^{2N}, \qquad \lambda = x\, d \left[\log\left(\frac{P_N(x)-y}{P_N(x) + y}\right)\right],
\end{equation}
with $\Lambda$ the dynamical scale related to the UV cutoff $\Lambda_{UV}$ and the UV coupling $\tau_{UV}$ as 
\begin{align}
    \Lambda^{2N}=\Lambda_{UV}^{2N} e^{2\pi i\, \tau_{UV}} \,.
\end{align}

\item The action $\Gamma: \mathcal M_{\text{CB}} \to \mathcal M_{\text{CB}}$ of the outer automorphism on the Coulomb branch parameters follows in a straightforward way from the analysis of the lagrangian in the previous section. We define
    \begin{equation}
    \label{eq:cbparametersdef}
    u_{n} \equiv \text{Tr}\left(\langle \Phi^n\rangle\right), \qquad n=2,\dots,N\,,
    \end{equation}
with $\Phi$ the adjoint scalar in the $\mathcal N = 2$ vector multiplet. More explicitly, this field transforms in the adjoint representation of the Lie algebra of $SU(N)$. Therefore, the action of the $\mathbb{Z}_2$ is the tangent map to the action on the Lie group \eqref{eq:homomorphism_semidirect},
\begin{align}
    \Phi \to -\Phi^T\,.
\end{align}
As a consequence, on the Coulomb branch, we have that
    \begin{equation}        \label{eq:explicitactiongamma}
        u_{n} \to (-1)^n u_{n}\,.
    \end{equation}
\end{itemize}

In what follows, we expect  to be able to lift the action of $\Gamma$ on the $\mathcal M_{\text{CB}}$ to an action $\widetilde{\Gamma}$ on the total space $\mathcal F$ of the SW geometry: 
  \begin{equation} \label{eq:liftingquotient}
  \begin{tikzcd}[ampersand replacement=\&]
      \mathcal F  \arrow[r,"/\widetilde{\Gamma}"] \arrow[d,"\pi"] \&  \widetilde{\mathcal F}  \arrow[d,"\widetilde{\pi}"]  \\
       \mathcal M_{\text{CB}} \arrow[r, "/\Gamma"]\& \widetilde{\mathcal M_{\text{CB}}}\\
    \end{tikzcd}\,,
    \end{equation}
where $\widetilde{\mathcal F}$ and $\widetilde{\mathcal M_{\text{CB}}}$ denote, respectively, the total space of the SW geometry and the Coulomb branch of the theory \textit{after} the discrete gauging; the horizontal arrows are the orbifold maps on $\mathcal F$ and $\mathcal M_{\text{CB}}$; and the vertical arrows $\pi,\, \widetilde{\pi}$ are the projection of the SW families on the Coulomb branches.

For $\Gamma$ to be a (spontaneously broken) symmetry on the CB, it must preserve the special K\"ahler metric. Let's see the consequences of this statement. Let 
\begin{equation}
\label{eq:specialcordsdef}
a_{i}(u) \equiv \int_{A_{i}(u)}\lambda(u), \qquad a_{D,i}(u) \equiv \int_{B_{i}(u)}\lambda(u),\qquad i = 1, ..., N-1
\end{equation}
be the ``special coordinates'' on the CB, with $u$ denoting a point in the CB and with $\left\{A_i(u),B_i(u)\right\}_{i=1}^{N-1}$ being a basis of $H_1(\pi^{-1}(u),\mathbb Z)$ with intersection product 
\begin{align}\label{eq:intersection_A_and_B_cycles}
    \langle A_{i}, A_j\rangle = 0, \quad \langle B_i, B_j \rangle= 0, \quad \langle A_i, B_j \rangle = \delta_{ij}\,.
\end{align}

We can locally invert the expressions $a_i(u)$ to obtain $u(a_i)$ and use $a_i$ as (local) coordinates on the CB. In these coordinates, the special K\"ahler metric reads $ds^2 = \tau_{ij}\, da^i\, da^j_D$, with 
\begin{equation}
    \label{eq:metricandspecialcords}
 \tau_{ij} = \frac{\partial a_{D,j}\rvert_{u = u(a)}}{\partial a_i}\,, \quad i,j = 1,...,N-1
\end{equation}
being related to the effective couplings of the $U(1)^r$ theory on the CB. 

Requiring that $\Gamma$ is an isometry of the CB therefore requires that the periods \eqref{eq:specialcordsdef} are preserved by its action. By Torelli's theorem, this is shown to imply that the fibers above $u$ and $\Gamma(u)$ must be isomorphic,
\begin{align}
  \pi^{-1}(u) \cong \pi^{-1}\left(\Gamma (u)\right)  \,,
\end{align}
and that $\Gamma$ lifts to an automorphism $\widetilde{\Gamma}$ of $\mathcal F$ (as in \eqref{eq:liftingquotient}). 

In fact, the action of $\widetilde{\Gamma}$ on the fibral coordinates $x,y$ must be non-trivial. This can be understood as follows: if we are in a mass-deformed phase of the SCFT point, $\Gamma$ will, in general, also exchange points where different BPS states have vanishing mass. If the two states are supported, respectively, on $\gamma \in H^1(\pi^{-1}(u),\mathbb Z)$ and $\gamma' \in H^1(\pi^{-1}(\Gamma(u)),\mathbb Z)$ (where $\gamma \neq \gamma'$) then the fibral action of $\widetilde{\Gamma}$ identifies $\gamma$ with $\gamma'$, namely
\begin{equation}
    \label{eq:matchingbpsstates}
    \widetilde{\Gamma}_{\ast}\gamma = \gamma',
\end{equation}
 with $\widetilde{\Gamma}_{\ast}: H_1(\pi^{-1}(u),\mathbb Z) \to H_1(\pi^{-1}\left(\Gamma(u)\right),\mathbb Z)$ being the action induced by $\widetilde{\Gamma}$ on the degree-one homologies.

As a final comment, we notice that, if the parent theory is conformal, in order to preserve $\mathcal N = 2$ supersymmetry, the $\widetilde{\Gamma}$ action has to commute with the $U(1)_R$ $R$-symmetry. In the context of SW geometry, the $U(1)_R$ $R$-symmetry gets promoted to a quasi-homogeneous $\mathbb C^*$ action, that acts with weights $[u_n]$ on the CB coordinates and with weights $[x], [y]$ on the fibral coordinates. Asking $\widetilde{\Gamma}$ to commute with  $\mathbb C^*$ hence, means that $[\Gamma(\cdot)] = [\cdot]$. This strongly restricts the possible actions in the fibral coordinates. In general, a biholomorphic change of coordinates on $\mathbb C^3 \ni (x,y,u)$ looks like
\begin{equation}
    \label{eq:generalcoordchange}
    \widetilde{\Gamma}(x) = c x + f_x(y,u),
\end{equation}
with $c$ a complex constant, and similarly for $y$. One can easily see that there are no polynomial combinations of $y,u$ that have the same $U(1)_R$ charge of $x$ and that can then appear at the r.h.s. of \eqref{eq:generalcoordchange} if we ask $\widetilde{\Gamma}(x)$ to have a well-defined $U(1)_R$ charge (and the same argument holds for $\widetilde{\Gamma}(y)$). Finally, recall that the outer automorphism is a $\mathbb{Z}_2$ symmetry, meaning that the action of $\widetilde{\Gamma}$ must square to 1. This further restricts the constant $c$ to be $\pm 1$ (independently for $x$ and for $y$). In the end, we conclude that
\begin{align}\label{eq:theformoftheaciton}
    \widetilde{\Gamma}(x,y,u_n) = (\pm x, \pm y, (-1)^nu_n)\,.
\end{align}

Note that whilst our argument requires the theory to be conformal (so that the $R$-symmetry is not anomalous), the form of the action \eqref{eq:theformoftheaciton} should also hold for the non-conformal case.  This is because one can introduce masses for the hypermultiplets and integrate them out without disturbing the discrete action.

Let's now proceed to compute the SW solution for $\widetilde{SU}(N)$ in explicit examples. The task amounts to, firstly, extract the $\widetilde{\Gamma}$ action in different cases; and secondly, compute the SW data of the ``daughter'' theory by taking the quotient, namely 
\begin{itemize}
    \item The new SW geometry, $\widetilde{\mathcal F} \equiv \mathcal F/\widetilde{\Gamma}$, with canonical projection $\widetilde{\pi}$ on the CB induced by the one of $\mathcal F$ \eqref{eq:liftingquotient}.
    \item The new SW differential $\widetilde{\lambda}$ obtained expressing $\lambda$ in terms of the $\widetilde{\Gamma}$-invariant coordinates. 
\end{itemize}

\subsection{\texorpdfstring{$\widetilde{SU}(3)$}{} Super Yang-Mills}
\label{sec:SW_su3}
We begin with the example of pure $\widetilde{SU}(3)$ SYM. The starting SW data of the parent theory are the particularization of  \eqref{eq:swcurvesunsymgeneral},
\begin{equation}
\label{eq:SWdatasu3sym}
    y^2 = P_3^2(x)  - 4 \Lambda^6, \qquad \lambda = x\, d \left[\log\left(\frac{P_3(x)-y}{P_3(x)+y}\right)\right] \,,
\end{equation}
with 
\begin{align}
   P_3(x) = x^3 + u_2 x + u_3 
\end{align}
and $\Lambda$ the dynamical scale. The action of $\Gamma$ on the CB parameters is
\begin{align}
    \Gamma(u_2,u_3) = (u_2, -u_3)\,.
\end{align}
There is only one choice that preserves the SW data \eqref{eq:SWdatasu3sym},
\begin{align}
    \widetilde{\Gamma}(x,y,u_2,u_3) = (-x,y,u_2,-u_3)\,,
\end{align}
such that $P_3$ picks up a sign that ends up cancelling in the SW curve and differential.

To compute the quotient, we introduce the invariant coordinates
\begin{equation}
    \label{eq:invcordssu3sym}
    a \equiv x^2, \quad b \equiv x u_3, \quad u_6 \equiv u_3^2\,.
\end{equation}
and their relation
\begin{equation}
    \label{eq:quotienteqsu3sym}
    a u_6 - b^2 = 0\,.
\end{equation}
We now have to express the SW data \eqref{eq:swdatasu3tildasym} in terms of the invariant coordinates. The SW curve, expressed in terms of the invariant coordinates, reads
\begin{equation}
\label{eq:finalresauxsu3sym}
y^2=\left(a+u_2\right) \left(a \left(a+u_2\right)+2 b\right)-4 \Lambda ^6+u_6\,.
\end{equation}
Equations \eqref{eq:finalresauxsu3sym} and \eqref{eq:quotienteqsu3sym} together define the full SW geometry for $\widetilde{SU}(3)$ as a complete intersection inside $\mathbb C^5 \ni (y,a,b,u_2,u_6)$, with the canonical projection on the CB being realized as the projection on $(u_2,u_6)$. Note that it is very important to remember the second equation, as \eqref{eq:finalresauxsu3sym} alone looks like a genus one curve at first glance.

In order to compute the SW differential, we can use that, for the $SU(3)$ SYM case \cite{Klemm:1995wp}, 
\begin{equation}
    \label{eq:differentialsu3sym}
    \lambda = \frac{x}{y} d P_3 = \frac{3x^2+u_2}{y} x dx= \frac{(3 a + u_2)}{2y} da\,,
\end{equation}
and hence can be promoted to a Seiberg-Witten differential for  the discrete-gauged curve:
\begin{equation}
    \label{eq:finalresultsu3symdifferential}
    \widetilde{\lambda} =  \frac{(3 a + u_2)}{2y} da\,.
\end{equation}
We can massage the result, solving \eqref{eq:quotienteqsu3sym}  for $a$ and plugging the solution into \eqref{eq:finalresauxsu3sym}, obtaining 
\begin{equation}
    \label{eq:finalresultsu3symwithpoles}
    y^2 = \frac{\left(b^3+b u_2 u_6+u_6^2\right)^2}{u_6^3}- 4 \Lambda ^6,
\end{equation}
where we can now easily see that for $u_6\neq 0$ we have a genus two curve as expected. We note that \eqref{eq:finalresultsu3symwithpoles} is \emph{not} in the canonical form $y^2 = \mathcal P(x,u_j)$, with $\mathcal P(x,u_j)$ a polynomial in $(x,u_j)$, but displays some poles at $u_6 = 0$. For the pure $SU(3)$ theory (and for its daughter theory $\widetilde{SU}(3)$) this might not seem a big issue: we indeed also have a point at infinity in the CB where other CB coefficients diverge. However, as we will see in  \cref{sec:SW_su36fl}, this will also happen in the conformal case, where there is no singularity at infinity. Indeed, the pole signals the fact that the true SW geometry is a complete intersection given by two equations in $\mathbb C^5$, not by one equation in $\mathbb C^4$. Consequently, we can only impose $a = \frac{b^2}{u_6}$ in the patch $u_6 \neq 0$ from \eqref{eq:quotienteqsu3sym}.

\subsection{\texorpdfstring{$\widetilde{SU}(3)+ 6 \, (F\oplus\overline{F})$}{}} \label{sec:SW_su36fl}

Let's consider now the superconformal case. The SW data, setting to zero the mass parameters $\mu$ and $m_i$ in \eqref{eq:swcurvesunfflavorsconf} (for $N=3$), leads to\footnote{
The SW curve for $SU(3)$ with six fundamental flavors is often written \cite{Argyres:2005pp} in the form,
\begin{equation*}
\label{eq:SWdatasu3martone}
    y^2 + x^6 + (v + u x)^2  + x^3(v + u x)\tau, \qquad \lambda = v \frac{dx}{y} + u \frac{x dx}{y}. 
\end{equation*}
One can match the notation of \cite{Argyres:2005pp} with ours by the following change of coordinates: 
\begin{equation}
    v_{\text{there}} = i u_3, \quad u_{\text{there}} = \frac{i u_2}{(f-1)^{1/6}}, \quad \tau_{\text{there}} = \frac{2 i}{\sqrt{f-1}},\quad  y_{\text{there}} = y, \quad x_{\text{there}} = (f-1)^{1/6} x\nonumber
\end{equation}}
\begin{equation}
\label{eq:SWdatasu3}
    y^2 = P_3^2(x) - f x^6, \quad \lambda = x d \Bigg[\log\left(\frac{P_3(x) - y}{P_3(x)+y}\right)\Bigg],
\end{equation}
with again $P_3(x) = x^3 + u_2 x + u_3$. Once again, there is only one choice of $\widetilde{\Gamma}$ that can preserve the SW data \eqref{eq:SWdatasu3},
\begin{equation}
    \label{eq:quotientautsu3}
    \widetilde{\Gamma}(x,y,u_2,u_3) = (-x,y,u_2,-u_3) \,.
\end{equation}

We note an interesting difference between the rank 1 case \cite{Argyres:2016yzz} (which we recap in appendix \ref{sec:SW_su2}) and our case. Here, \eqref{eq:quotientautsu3} is an automorphism of the SW geometry \eqref{eq:SWdatasu3} even without imposing any tuning on the UV coupling $\tau$. Indeed, the outer automorphism of the gauge group is a symmetry of the theory for all the values of the UV coupling, rather than a duality that gets enhanced to a symmetry for specific values of $\tau$ (as in the case of $SU(2))$.

Now we proceed as in the pure SYM case. The invariant coordinates are 
\begin{equation}
    \label{eq:invcordssu3}
    a \equiv x^2, \quad b \equiv x u_3, \quad u_6 \equiv u_3^2\,,
\end{equation}
and satisfy
\begin{equation}
    \label{eq:quotienteqsu3}
    a u_6 - b^2 = 0.
\end{equation}
We then express the SW geometry in terms of \eqref{eq:invcordssu3}, obtaining
\begin{equation}
    \label{eq:quotientswequationsu3}
    y^2 = -4 a^3 f+a^3+2 u_2 \left(a^2+b\right)+2 a
   b+a u_2^2+u_6.
\end{equation}
Equations \eqref{eq:quotientswequationsu3} and \eqref{eq:quotienteqsu3} define the SW geometry of the daughter theory, with the canonical projection on the Coulomb Branch base being the projection over $(u_2, u_6)$. 

We can again massage the expression for the SW geometry. We proceed first by solving \eqref{eq:quotienteqsu3} to get $a = \frac{b^2}{u_6}$, and then we plug it in \eqref{eq:quotientswequationsu3},
\begin{equation}
    \label{eq:finalresauxsu3}
    y^2 = \frac{b^6 (1-4 f)+2 b^4 u_2 u_6+b^2 u_6^2
   \left(2 b+u_2^2\right)+2 b u_2
   u_6^3+u_6^4}{u_6^3}.
\end{equation}
Let us now concentrate on the SW differential. In the parent theory, we can rewrite it as
\begin{equation}
    \label{eq:swparentsu3conf}
    \lambda= \frac{2 u_2 x + 3 u_3}{y}dx \,.
\end{equation}
This is already, up to numerical constants, in the canonical form of \cite{Argyres:2005pp}. We now want to obtain $\widetilde{\lambda}$, by expressing $\lambda$ in terms of invariant coordinates. One can check that 
\begin{equation}
\label{eq:lambdatildasu3conf}
\widetilde{\lambda} = \lambda\left(x= x(a,b,u_6),u_3 = u_3(a,b,u_6)\right) = \frac{u_2 + \frac{3 b}{2 a}}{y}da\,.
\end{equation}
We can massage \eqref{eq:lambdatildasu3conf} as well, imposing $a = \frac{b^2}{u_6}$, obtaining 
\begin{equation}
    \label{eq:lambdatildesu3confsimpler}
    \widetilde{\lambda} = \frac{2  u_2 }{u_6} \frac{b}{y} db+\frac{3}{y} db\,.
\end{equation}
We want now to express the result in coordinates where the SW differential \cite{Argyres:2005pp} is $\lambda = u_2 X d b/Y + u_6 d b/Y$, with the curve in the form $u_6^k Y^2 + P(b,u) = 0$. This can be achieved uniquely by taking $ y = \frac{Y}{u_6}$. In these coordinates, the curve takes the nicer, yet not ``polynomial'' form
\begin{equation}
\label{eq:finalressu3}
    Y^2 =    \frac{b^6 (1-4 f)+2 b^4 u_2 u_6+b^2 u_6^2
   \left(2 b+u_2^2\right)+2 b u_2
   u_6^3+u_6^4}{u_6}\,.
\end{equation}
This pole is responsible for the new singularity in the CB after the discrete gauging that we advertised in the introduction, and which we'll discuss at lenght in Section \ref{sec:bps_quiver}.

It turns out that the presence of this coefficient is good news. In the classification of Argyres and Martone \cite{Argyres:2022lah} there is no theory with rank 3 global symmetry (as we expect from the lagrangian analysis from Section \ref{sec:lagrangian}; and will check momentarily in the curve) and CB parameters of scaling dimension $\left\{2,6\right\}$. But that reference restricts to geometries such that the coefficients $f_i(u)$ in the r.h.s. of the SW curve in canonical form are polynomials of the CB parameters. This coefficient, leading to a pole in the r.h.s., is what allows us to escape the 
the classification of \cite{Argyres:2022lah}.  We hence see that the discrete gauging of charge conjugation in $SU(3) + 6$ flavors is \textit{a new} rank-two SCFT beyond the classification of \cite{Argyres:2022lah}, and more generally, that there can be rank 2 SCFTs with non-polynomial coefficients.

Let's now compute the number of flavor parameters of the discrete-gauged theory from the IR. The mass-deformed curve before the discrete gauging is  \eqref{eq:swcurvesunfflavorsconf}, which we recall here for convenience: 
\begin{equation}
\label{eq:swcurvemassessu3}
y^2 = P_3^2(x) - f \prod_{i=1}^6(x-m_i)\,.
\end{equation}

We then see that requiring that \eqref{eq:quotientautsu3} is a symmetry selects just the even terms in the product $\prod_{i=1}^6(x-m_i)$, and we are left with three mass deformations. In particular, the degree-one mass Casimir invariant $\mu = \frac{1}{6}\sum_{i=1}^6 m_i$ has to be set to zero, which is the IR avatar of the fact that the $\mathbb{Z}_{2}$ outer automorphism acts non-trivially on the $\mathfrak{u}(1)$ center factor of the flavor group. It is important to remark that $\mu \to 0$ is also important to get an invariant SW differential in the mass-deformed theory.  We can understand better the final flavor group by noticing that setting to zero  $\sum_{i=1}^6 m_i$ reduces first the flavor group from $\mathfrak{u}(6) \to \mathfrak{su}(6)$. Then, setting to zero the odd-degree Casimirs of $\mathfrak{su}(6) \cong A_5$ leads us to a Lie algebra 
compatible with $C_3$ (or $D_3$). This is precisely the flavour symmetry we expected from the lagrangian, which constitutes one of the checks of our computation.

\subsection{\texorpdfstring{$\widetilde{SU}(4)+ 8 \, (F\oplus\overline{F})$}{}}\label{sec:SW_su48fl}

Let's now increase the rank and compute the SW curve for the SCFT with gauge group $\widetilde{SU}(4)$. The SW curve and differential of the parent theory are
\begin{align} \label{eq:SWsu4}
        y^2 = P_4^2(x) - f x^8, \quad \lambda = x \, d\Bigg[\log \left(\frac{P_4(x) - y}{P_4(x)+y}\right)\Bigg]\,,
\end{align}
where
\begin{align}
    P_4(x) = x^4 + u_2 x^2 + u_3 x + u_4\,.
\end{align}

The action of the $\mathbb{Z}_2$ on the base is
\begin{align}
    \Gamma (u_2,u_3,u_4) = (u_2,-u_3,u_4) \,.
\end{align}
We notice that the action of the fibral coordinates cannot be the same as in $\widetilde{SU}(3)$. Flipping the sign of the coordinate $x$ now leads to $P_4$ being invariant and the SW differential getting an overall sign. In order to compensate it, also $y$ needs to pick a sign so that \eqref{eq:SWsu4} is preserved,
\begin{align}\label{eq:gammasu4}
    \widetilde{\Gamma} (x,y,u_2,u_3,u_4) = (-x,-y,u_2,-u_3,u_4)\,.
\end{align}

From this point we proceed analogously to the rank 2 case. The ring of invariant coordinates has generators,
\begin{eqnarray}
    \label{eq:invcordssu4}
    a_1 &\equiv & x^2, \quad a_2 \equiv y^2, \quad a_3 \equiv x y, \nonumber \\ 
    u_6 &\equiv & u_3^2, \quad b_1 \equiv x u_3, \quad b_2 \equiv y u_3\,. 
\end{eqnarray}
To find all the relations between them, we first build the matrix
\begin{equation}
    \label{eq:quotienteqsu4}
\mathcal M =\left(
 \begin{array}{ccc}
 a_1 & a_3 & b_1 \\
 a_3 & a_2 & b_2 \\
 b_1 & b_2 & u_{6} \\
\end{array}
\right)\,,
\end{equation}
and then set all its $ 2 \times 2 $ minors equal to zero,
\begin{equation}
    \label{eq:minorseqsu4}
    \text{Min}_{2 \times 2}(\mathcal M) = 0\,.
\end{equation}
We also have the equation \eqref{eq:SWsu4}, which in terms of \eqref{eq:invcordssu4} becomes
\begin{align}\label{eq:curvesu4tilde}
    a_2 + a_1^4 (f-1) = 2 a_1^3 u_2 + 2 a_1 u_2 (b_1+u_4) + (b_1+u_4)^2 + a_1^2 (2 b_1+u_2^2+2 u_4)\,.
\end{align}
Together with \eqref{eq:minorseqsu4}, \eqref{eq:curvesu4tilde} defines the SW geometry of $\widetilde{SU}(4)$.  We note that the SW geometry is of complex dimension four, and is not a complete intersection inside the ambient space $\mathbb C^8 \ni (a_1,a_2,a_3,b_1,b_2,u_6,u_2,u_4)$ since the number of equations that identify it is bigger than its codimension in $\mathbb C^8$.   

Lastly, the SW differential before the discrete gauging can be put into the following form,
\begin{align}
    \lambda = -\left( 8 u_4 + 6 u_3 x + 4 x^2 u_2 \right) \frac{dx}{y}\,.
\end{align}
Note that, as expected, all the terms are invariant under $\widetilde{\Gamma}$ \eqref{eq:gammasu4}. Therefore,
\begin{align}
    \widetilde{\lambda} = -\left(4 u_4 + 3 b_1 + 2 u_2 a_1\right) \frac{d a_1}{a_3}\,,
\end{align}
where we have used that, from the relations \eqref{eq:minorseqsu4}, $dx/y=d a_1/2a_3$.

We finish with three remarks. First, note that while in this case we have not given the expression for the curve after trying to plug the solutions of \eqref{eq:minorseqsu4} into \eqref{eq:curvesu4tilde}, it is clear that doing so would lead to pole singularities, similarly to the $\widetilde{SU}(3)$ case. Second, notice that the lift of the action of the $\mathbb{Z}_2$ to the full SW fibration in this case includes flipping also the sign of the coordinate $y$, which results in the quotient geometry involving several more generators and relations. As mentioned in the introduction, this is one of the reasons that make difficult obtaining a closed form expression for the SW solution of $\widetilde{SU}(N)$ for generic $N$. It is easy to convince oneself that the different action on the coordinate $y$ will depend on whether $N$ is even or odd. And last, we see that, as in the $\widetilde{SU}(3)$ case, the flip of the sign of the coordinate $x$ prevents us from having mass deformations corresponding to odd-degree Casimirs of the flavour algebra. This is compatible with either the $C_4$ or $D_4$ subalgebras of the original $A_7$, which is consistent with the flavour symmetry of the lagrangian $USp(8)$.

\subsection{\texorpdfstring{$\widetilde{SU}(5)+ 10 \, (F\oplus\overline{F})$}{}}
\label{sec:SW_su510fl}
Next, we will repeat the analysis we just outlined for $SU(4)$ and $SU(3)$ groups for the $SU(5) + 10 F$ theory. This case is particularly interesting since the CB of the discretely gauged theory is \textit{not} freely generated and displays complex structure singularities. Indeed, the CB parameters of the parent theory are the Casimir invariants $u_2, u_3, u_4, u_5$ of the UV theory adjoint complex scalar. The discrete gauging leaves even-degrees Casimir invariant, while it acts on $u_3$ and $u_5$ as 
\begin{equation}
    \label{eq:SU(5)actiononCB}
    u_3 \to - u_{3}, \qquad u_5 \to - u_5.
\end{equation}
The new CB, after the discrete gauging, is hence described by five invariant coordinates $u_2,u_4,u_6 \equiv u_3^2, u_{10} \equiv u_5^2, u_{8} \equiv u_3 u_5$ satisfying
\begin{equation}
    \label{eq:invariantcordsSU(5)}
    u_{10} u_6  - u_{8}^2 = 0\,.
\end{equation}
We see that \eqref{eq:invariantcordsSU(5)} is the equation of an $A_1$ Du Val singularity, and the CB of the daughter theory is, as complex manifold, $\mathbb C^2 \times A_1$, where $u_2,u_4$ span the $\mathbb C^2$ factor. 

We now want to obtain the SW solution for the discrete gauging of the charge conjugation of $SU(5) + 10 F$. We will proceed as in the previous cases, first lifting (uniquely) the charge conjugation action to the SW geometry of the parent theory $\mathcal F$, and then performing the quotient to obtain $\widetilde{\mathcal F}$. The SW solution for $SU(5) + 10 F$ is
\begin{equation}
    \label{eq:SWdataSU(5)}
    y^2 = P_5^2(x)  - f x^{10}, \quad \lambda = x d \left[\log\left(\frac{P_5(x)-y}{P_5(x) + y}\right)\right]\,,
\end{equation}
with
\begin{align}
    P_5(x) = x^5 + u_2 x^3 + u_3 x^2 + u_4 x + u_5.    
\end{align}

The only possible lift of the action $\Gamma$ to $\mathcal F$ is 
\begin{equation}
    \label{eq:quotientautsu5}
\widetilde{\Gamma} (x,y,u_2,u_3,u_4,u_5) = (-x,y,u_2,-u_3,u_4,-u_5)\,.
\end{equation}
The ring of invariant coordinates is generated by
\begin{eqnarray}
    \label{eq:invcordssu5}
    a_1 &\equiv & x^2, \quad b_1 \equiv x u_3, \quad b_2 \equiv x u_5, \nonumber \\ 
    u_6 &\equiv & u_3^2, \quad u_{10} \equiv u_5^{2}, \quad u_8 \equiv u_3 u_5. 
\end{eqnarray}
The relations satisfied by \eqref{eq:invcordssu5} are found by first constructing the following matrix,
\begin{equation}
    \label{eq:quotienteqsu5}
\mathcal M =\left(
 \begin{array}{ccc}
 a_1 & b_1 & b_2 \\
 b_1 & u_6 & u_8 \\
 b_2 & u_8 & u_{10} \\
\end{array}
\right)\,,
\end{equation}
and then setting to zero its two by two minors,
\begin{equation}
    \label{eq:minorseq}
    \text{Min}_{2 \times 2}(\mathcal M) = 0.
\end{equation}
We note that \eqref{eq:invariantcordsSU(5)} corresponds to the relation associated to the minor obtained by removing the first row and the first column of \eqref{eq:quotienteqsu5}.

As in the previous cases, the next step is to express the SW geometry $\mathcal F$ in terms of the invariants coordinates \eqref{eq:invcordssu5}: 
\begin{align}
    \label{eq:quotientswequationsu5}
y^2 =& a_1^3 \left(2 b_1+u_2^2+2 u_4\right)+a_1^2
   \left(2 u_2 \left(b_1+u_4\right)+2
   b_2+u_6\right)\nonumber\\
   &+a_1 \left(2 b_1 u_4+2 b_2
   u_2+u_4^2+2 u_8\right)+a_1^5 (-(f-1))+2
   a_1^4 u_2+2 b_2 u_4+u_{10}
\end{align}
The equation \eqref{eq:quotientswequationsu5}, together with \eqref{eq:minorseq}, describes the SW geometry $\widetilde{\mathcal F}$ of the daughter theory $\widetilde{SU}(5) + 10 (F\oplus\overline{F})$. The projection on the Coulomb branch basis $\widetilde{\pi}: \widetilde{\mathcal F} \twoheadrightarrow \widetilde{\mathcal M_{\text{CB}}}$  is realized as the projection on the $u_{n}$ coordinates, that satisfy the non-trivial relation \eqref{eq:invariantcordsSU(5)}. 
If we restrict ourselves to a Zariski-open subset of $\widetilde{\mathcal M_{\text{CB}}}$, we can obtain an expression for the SW geometry that explicitly shows that the genus of the SW curve is four. In the $u_6 \neq 0$ open set of $\widetilde{\mathcal M_{\text{CB}}}$, we find 
\begin{eqnarray}
    \label{eq:finalresauxsu5}
y^2 &=&\frac{b_1^{10} (1-f)+2 b_1^8 u_2 u_6+2
   b_1^7 u_6^2+b_1^6 \left(u_2^2+2
   u_4\right) u_6^2+2 b_1^5 u_2 u_6^3}{u_{6}^5}+ \nonumber \\
   &&\frac{b_1^4
   u_6^3 \left(2 b_2+2 u_2 u_4+u_6\right)+2
   b_1^3 u_4 u_6^4+b_1^2 u_6^4 \left(2 b_2
   u_2+u_4^2+2 u_8\right)}{u_6^5}+2 b_2
   u_4+u_{10} \nonumber \\
\end{eqnarray}
with the Coulomb branch parameters $u_n$ satisfying \eqref{eq:invariantcordsSU(5)}.

Let's now concentrate on the SW differential: one can rewrite it, using that $y^2 = P_5^2 - f x^{10}$ and hence $dy = d(P_5^2 + f x^{10})/2 y$, as  
\begin{equation}
\label{eq:lambdaswrewrittensu10}
    \lambda =2 (-2 x^5 + x^2 u_3 + 2 x u_4 + 3 u_5)  \frac{dx}{y} \,.
\end{equation}
Once again, all the terms appearing in the rightmost side of \eqref{eq:lambdaswrewrittensu10} are explicitly invariant under $\widetilde{\Gamma}$. Consequently, we can easily re-express $\lambda$ in terms of the invariant coordinates \eqref{eq:invcordssu5} obtaining in this way the SW differential $\widetilde{\lambda}$ of the daughter theory:
\begin{equation}
    \label{eq:swdifferentialdaughtersu5}
    \widetilde{\lambda} =  \left(
   2 a_1 u_2+3 b_1+4
   u_4+5
   b_2/a_1 \right) da_1\,.
\end{equation}
All together, \eqref{eq:swdifferentialdaughtersu5}, \eqref{eq:quotienteqsu5} and \eqref{eq:quotientswequationsu5} define the SW solution of the discrete gauging of the outer automorphism of $SU(5) + 10$ flavors. This is the first explicit example of a SW solution for a 4d $\mathcal N = 2$ theory with non-freely generated Coulomb branch!

We can conclude this section by performing, as in the previous cases, the counting of the mass parameters of the daughter theory. The mass-deformed curve of $SU(5) + 10 F$ is 
\begin{equation}
\label{eq:swcurvemassessu5}
    y^2 = P_5^2(x) - f\prod_{i=1}^{10}(x-m_i) = P_5^2(x) + Q(x,m).
\end{equation}
We see from \eqref{eq:swcurvemassessu5} that $\widetilde{\Gamma}$ extends to an automorphism on the of the extended Coulomb branch when 
$Q(-x,m) = Q(x,m)$. Consequently, we have to set to zero all the odd-degree mass Casimir invariants of the starting flavor symmetry, selecting in this way a $C_5$ (or $D_5$) subalgebra of the original $\mathfrak{su}(10) \cong A_9$ flavor algebra. This is, as in the previous cases, in agreement with the computation we performed in \cref{sec:lagrangian}.

\subsection{Higher ranks}
\label{sec:higherranks}
Even if we are not able to provide a closed form expression for $\widetilde{SU}(N)$, we can comment on the general structure of the discrete gauging procedure for generic rank. As we saw, the action on the coordinates $(x,y,u_n)$ can be extracted just with geometric methods, but we can, a posteriori, find a motivation for it. Recall the action of the outer automorphism on the UV adjoint scalar $\Phi \to - \Phi^T$ and the generic form of the SW solution for $SU(N)  + N_f F$ \cite{Argyres:1995wt},
\begin{equation}
    \label{eq:swcurvesunfflavorsgeneral}
    y^2 = P_N^2(x) - 4 \Lambda^{2N - N_f}\prod_{i=1}^{N_f}(x-m_i), \qquad \lambda = x\, d \left[\log\left(\frac{P_N(x)-y}{P_N(x) + y}\right)\right]\,,
\end{equation}
with $P_N(x) = \text{det}(\mathds{1}x - \Phi)$ and \eqref{eq:swcurvesunfflavorsconf} for the conformal case. Consequently, since \cite{DeMarco:2022dgh} $x$ has to transform as\footnote{It would be interesting to apply our geometric techniques to the 5d discrete gauging setup of \cite{Collinucci:2021ofd,Collinucci:2022rii,DeMarco:2021try,DeMarco:2022dgh}. In that case, we are interested in the Higgs branch of the theory obtained by putting M-theory on a Calabi-Yau space. Similarly to our setup, the physics on the 5d moduli space of vacua can be understood in terms of the complex deformation of a geometric space (in that case the Calabi-Yau variety).} the eigenvalues of $\Phi$, and hence as $\Phi$,  we have $[x] = [\Phi]$, and hence we must always have $x \to - x$. Furthermore, we have $u_n \propto \langle \Phi^n\rangle$ and hence the discrete-gauging acts as $u_j \to (-1)^n u_n$. This leads to $P_N \to (-1)^N P_N$. By looking to \eqref{eq:swcurvesunfflavorsgeneral}, we see that the curve is already preserved by this action, but, to make the SW differential invariant w.r.t. to $\widetilde{\Gamma}$, we need to impose $y \to (-1)^{N+1} y$.

For all $N$, we then reflect at least two coordinates among $x,y, u_n$, and hence in general the final SW geometry \textit{is not} a hypersurface equation, with the Coulomb branch not freely generated and displaying complex structure singularities for $N \geq 5$.  Curiously, we notice that for $N \geq 7$, the Coulomb branch of the daughter theory can be non-Gorestein (e.g., isomorphic to $\mathbb C^3 \times \frac{\mathbb{C}^3}{\mathbb Z_2}$, with the $\mathbb Z_2$ acting as $- \mathds{1}$ on the coordinates) in the particular case of $SU(7)$.  

We conclude with a remark on the flavor symmetry: in all the cases, $x \to -x$ forces the odd-degree symmetric polynomials of the masses appearing in the expansion of $\prod_{i=1}^{N_f}(x-m_i)$ to be zero. If $N$ is odd, this automatically leads to a $C_{N_f/2}$ flavour symmetry. If $N$ is even, this criteria cannot be used to distinguish between the subalgebras $C_{N_f/2}$ or $D_{N_f/2}$. In principle, one could get more information by studying the explicit action of $\mathbb{Z}_2$ on the Weyl group of the flavor symmetry, but we postpone this analysis to future work. 

These differences between $N$ even and odd are reminiscent of the two possibilities for the discrete gauging of the outer automorphism for even $N$ studied in \cite{Arias-Tamargo:2019jyh}. The two disconnected gauge theories $\widetilde{SU}(2N)_{I}$ and $\widetilde{SU}(2N)_{II}$ have precisely a $USp(N_f)$ or $SO(N_f)$ global symmetry respectively. It would be interesting to understand if this apparent correspondence is coincidental or there is a deeper reason for it.

\section{BPS Spectra}\label{sec:bps_quiver}

In this section, we will comment on the BPS spectra of the analyzed theories, furnishing a complete computation in the $\widetilde{SU}(3)$ SYM case. We will start, in \cref{sec:swtospectrareview} briefly reviewing how to compute the generators of the BPS spectrum of a 4d $\mathcal N = 2$ theory from its SW solution. Then, in \cref{sec:generalbpsspectra}, we will list our general results for the BPS spectra of $\widetilde{SU}(3) + N_{f} (F \oplus \overline{F})$, discovering the existence of a new singularity $u_{fixed} \in \widetilde{\mathcal M_{\text{CB}}}$ at the points that are images of fixed points of  $\Gamma$. We will then explicitly check our results, in \cref{sec:bpsspectrumsu3tilde} computing the spectrum of $\widetilde{SU}(3)$ SYM. Finally, in \cref{sec:generalresultzkspectra}, with a general argument based upon the monodromies of the IR gauge coupling, we will predict the existence of singularities analogous to $u_{fixed}$ in the, up to mild assumptions, general situation of a discrete gauging of a $\mathbb Z_k$ symmetry of a generic, non necessarily lagrangian, 4d $\mathcal N = 2$ theory.

\subsection{BPS charges from the Seiberg-Witten solution}
\label{sec:swtospectrareview}

In Seiberg-Witten theory, a BPS particle is identified with particular elements $\nu$ of $ H_1(\pi^{-1}(u),\mathbb Z)$. Let's fix a symplectic system of generators $\left\{A_i,B_i\right\}_{i=1}^{r}$ of $H_{1}(\pi^{-1}(\vec{u}),\mathbb Z)$, satisfying \eqref{eq:intersection_A_and_B_cycles};
then, the central charge of the BPS particle associated with the cycle $\nu =  \sum_{i=1}^{r}g_iB_i + q_i A_i$ is 
\begin{equation}
    \label{eq:bpschargeformula}
    \mathcal Z(\nu) = \int_{\nu} \lambda = \sum_{i=1}^{r}g_ia_{D,i} + q_i a_i\,,
\end{equation}
with $g_i, q_i$ valued in $\mathbb Z$ being, respectively, the magnetic and electric charges of the particle. For a BPS particle, we have the condition 
\begin{equation}
    \label{eq:bpscentralchargevsmass}
    m(\nu) = \abs{Z(\nu)}\,,
\end{equation}
with $m(\nu)$ the mass of the particle. Consequently, the vanishing of $Z(\nu)$ selects a complex codimension-one locus on the Coulomb branch where the considered particle becomes massless. The union of all such loci coincides with the singular locus $\mathcal S$ of the Coulomb branch special metric, and $(a_{D,i},a_i)$ become bad coordinates when we approach $\mathcal S$. Physically, this is due to the fact that, along the naive RG flow, we have integrated out at least one particle whose mass, in an analytic neighborhood of the singularity, is comparable to the one of the Coulomb branch modes.

More precisely, let $\mathcal S_{j}$  be one of the irreducible components of $\mathcal S \subset \mathbb C^r$, where the particle associated to a  cycle $\nu_j \in H_1(\pi^{-1}(u),\mathbb Z)$ becomes massless. Let $\gamma_j \in \pi_{1}(\mathbb C^r \setminus \mathcal S)$ be a path that goes around $\mathcal S_j$ once, without linking with any other component $S_k \neq S_j$. Then, $(a_{D,i},a_i)$ develop the following monodromy along $\gamma_j$,
\begin{equation}
    \label{eq:monodromyfromchargesBIS}
    (a_{D,i},a_i)^T \longrightarrow M(\gamma_j) (a_{D,i},a_i)^T \,.   
\end{equation}
Here, we are picking a choice of convention, which matches the one of \cite{Klemm:1994qj}, such that the monodromy matrix acts on the charges from the right,
\begin{align}
    (g_i,q_i) \to (g_i,q_i)M(\gamma_j)\,,
\end{align}
where $M(\gamma_j) \in Sp(2r,\mathbb Z)$\footnote{Physically, the condition that $M(\gamma_j)$ lies in the symplectic group is due to the fact that \eqref{eq:monodromyfromcharges} realizes an electro-magnetic duality in the IR theory and the Dirac pairing
\begin{equation}
    \label{eq:diracpairing}
    \langle\nu_1,\nu_2\rangle \equiv (\vec{g}_1, \vec{q}_1)^T \left(
\begin{array}{cc}
 0 & \mathds{1} \\
 -\mathds{1} & 0 \\
\end{array}
\right) (\vec{g}_2, \vec{q}_2)
\end{equation}
must be preserved by such transformations.}. The monodromy matrix associated to a particle of magnetic and electric charges $(\vec{g},\vec{q}) = (g_i,q_i)$, with $i=1,...,r$, is \cite{Klemm:1994qj}  
\begin{equation}
    \label{eq:monodromyfromcharges}
    M(\nu) = \left(
\begin{array}{cc}
 \mathds{1}_{r}-\vec{q} \otimes \vec{g} & -\vec{q} \otimes\vec{q} \\
 \vec{g}\otimes \vec{g} & \mathds{1}_r + \vec{g} \otimes \vec{q}   \\
\end{array}
\right)\,.
\end{equation}
From the perspective of the SW geometry, \eqref{eq:monodromyfromcharges} is re-obtained by noticing that $(a_{D,i},a_i)$ are the periods of the Seiberg-Witten differential $\lambda$, that is a holomorphic one-form along $\pi^{-1}(u)$, varying holomorphically (hence without monodromies) on the Coulomb branch. Consequently, the monodromies of the $\lambda$ periods, as functions of the Coulomb branch coordinates $u$, must come from the monodromies of our choice of basis $\lbrace B_i(u),A_i(u)\rbrace$ on the Coulomb branch. The theory regulating the monodromies of a choice of basis $\lbrace B_i(u),A_i(u)\rbrace$ of the mid-dimensional homology in a family $\mathcal F$ of deformed varieties is well known \cite{arnold1985singularities2}, and is regulated by the so-called Picard-Lefschetz formula. 

Let us recall the Picard-Lefschetz formula for the case in which the fiber of the SW geometry is a complex curve. Let $\nu_j \in H_1(\pi^{-1}(u),\mathbb Z)$ be a one-cycle vanishing along a component $\mathcal S_j \subset \mathcal S \subset \mathbb C^r$. Then, for each $\beta \in H_1(\pi^{-1}(u),\mathbb Z)$, we have the following monodromy transformation going once around $\gamma_j$,
\begin{equation}
    \label{eq:picardlefschetz}
    \beta \to \beta  - \langle \beta,\nu \rangle \nu\,.
\end{equation}
Equation \eqref{eq:picardlefschetz} then induces  a linear action on $H_1(\pi^{-1}(u),\mathbb Z)$  that exactly coincides with the one of \eqref{eq:monodromyfromcharges}.

It might also be the case that, along $\mathcal S_j$, many BPS particles become massless at the same time. If this happens, the monodromy is $M(\mathcal S_j) = \prod_{k=1}^n M(U_{k})$, with $k = 1,..., n$ indexing the BPS particles simultaneously massless on $\mathcal S_j$. In this case, we can not use the Picard-Lefschetz formula \eqref{eq:picardlefschetz}, but we are still able, by carefully tracking the motion of the one-cycles of the curve (e.g. using the code of \cite{Bhardwaj:2021zrt}), to obtain the action $M(\mathcal S_j)$ on $H_1(\pi^{-1}(u),\mathbb Z)$. More explicitly, in our case all the SW curves are written in hyperelliptic form $y^2 = R_{2N}(x,u_j,m_{i})$, with $R_{2N}$ a polynomial in $x$ with coefficients depending on masses and Coulomb branch parameters. Consequently, the projection on the $x$ coordinates induces a double cover of the projective line $\mathbb P^1 \supset \mathbb C \ni x$ and the roots $x_k(u_j,m_i)$ of $R_{2N}$ are the branching points of the double cover. We can then construct the one-cycles of the curve by fibrering the solutions $y_{\pm} = \pm \sqrt{R_{2N}}$ over a real path joining two of the $x_k$. Therefore, we identify a vanishing cycle from requiring that (at least) a pair of roots $x_k$ coincide.

In practice, one picks a generic complex line $\mathcal L$ through a basepoint on the Coulomb branch (which is to say fixing a value for $u_j$) and then computes the monodromies around the points in $\mathcal L \cap \mathcal S$. Using the result of Van Kampen and Zariski (see e.g. \cite{Ceresole:1993nz}), if the slice is generic, we can then compute a system of generators for the BPS particles with the following procedure:

\begin{enumerate}
\item Identify the vanishing cycles on top of the considered point $u_{\text{sing}}\in\mathcal L \cap \mathcal S$ by looking to which branching points of the hyperelliptic curve double cover collide for $u = u_{\text{sing}}$. 

\item Compute the monodromy for the other cycles using the Picard-Lefschetz formula \eqref{eq:picardlefschetz}, or by brute force, e.g. using the code of \cite{Bhardwaj:2021zrt} to visualize the monodromies of the hyperelliptic double cover branching points.

    \item Compute the charges of the BPS particle becoming massless at the points in which $\mathcal L$ intersects $\mathcal S$ by comparing with \eqref{eq:monodromyfromcharges}.
    
    \item Select a system of generators (over $\mathbb Z$) out of the aforementioned charges. 
\end{enumerate}
The monodromies that found on $\mathcal L$ by this procedure must satisfy the following topological condition: 
\begin{equation}
    \label{eq:topologicalconditiongeneral}
    M_N...M_1=M_{loop},
\end{equation}
with $M_{loop}$ being the monodromy computed on a large anticlockwise loop inside $\mathcal L$ starting from some basepoint $u_{b}$. The matrices in the l.h.s. of \eqref{eq:topologicalconditiongeneral} are ordered according to the sequence in which they get touched by the anticlockwise rotation of a semi-infinite ray starting at the basepoint. This condition, which we have presented in the mathematical language (as it applies to any hyperelliptic curve, regardless of the context), has the simple physical interpretation that the monodromy picked up when going around all the singularities on the Coulomb branch has to match the monodromy at infinity computed from the 1-loop $\beta$-function.

\subsection{BPS spectra and charge conjugation: general results}
\label{sec:generalbpsspectra}
In this section, we gather the general results for the BPS spectra of the $\mathbb Z_2$ gauging of $SU(N) + F$ flavors, which we will later verify explicitly in the example of $SU(3)$ in \cref{sec:bpsspectrumsu3tilde}.

We remark that all the results in this section are based purely on the geometry of the SW solution. Hence, working with the assumption that the $\mathbb Z_2$ action on the Coulomb branch is accompanied by a $\mathbb Z_2$ action on the fiber, we believe that our results can be applied also to a more generic gauging of an abelian zero-form symmetry. We'll use this insight to extend our results to the case of a $\mathbb Z_k$ gauging of a generic (even non-lagrangian) 4d $\mathcal N  = 2$ theory in \cref{sec:generalresultzkspectra}.

Let $\mathcal L$ be a slice on the Coulomb branch $\mathcal M_{\text{CB}}$ of the parent theory. The summary of our results is as follows:
\begin{itemize}
\item For $SU(N) + F$ theories \cite{Klemm:1994qj,Klemm:1995wp}, if $(g_i,p_i)$ and $(g_i',p_i')$ are the charges of two particles becoming massless at two points $p, p'= \Gamma(p)$ of the parent theory Coulomb branch, then 
\begin{equation}
    \label{eq:electromagdualgeneral}
    M_p = V_N M_{p'} V_N, \qquad V_N^2 = \mathbbm{1}_{2r}
\end{equation}
with $V_N$ the electromagnetic duality realizing the $\mathbb Z_2$ action on the parent theory at the level of the SW fiber.
    \item We expect that, if a BPS particle becomes massless at a point $\widetilde{p} \in \widetilde{\mathcal M_{\text{CB}}}$ whose preimages are $p, p' = \Gamma(p)$, then either 
    \begin{equation}
    \label{eq:bpsspectragenres1}
    M_{p} = M_{\widetilde{p}}  \quad \text{or} \quad      M_{\widetilde{p}} = M_{p'} = V_N M_{\widetilde{p}} V_N,
    \end{equation}
    where, for the rightmost equality, we used \eqref{eq:electromagdualgeneral}. 
This is because going once around $\widetilde{p}$ corresponds, as observed in \cite{Argyres:2016yzz}, to go \textit{once} around, say $p$. Consequently, the monodromies are identified up to the action of the electromagnetic duality $V_{N}$ (that signals the fact that we could have picked $p'$ instead of $p$ as the preimage of $\widetilde{p}$). 
    \item If the considered slice $\mathcal L$ is an invariant subset of the parent theory Coulomb branch under the $\mathbb Z_2$ action, from the 1-loop computation we must have
    \begin{equation} 
\label{eq:monodromyconditiongeneral}
    M_{loop, daughter}^2 = M_{loop,parent}.
    \end{equation} 
as a consequence of the squaring of the Coulomb branch coordinates.

    \item If $u_0 \in \mathcal M_{\text{CB}}$ is a fixed point of the $\mathbb{Z}_2$ action, it gains a monodromy after the discrete gauging even if there is no BPS particle becoming massless at $u_0$ in the parent. The monodromy is given by $M_{\text{fixed}} = V_N$.\footnote{At this point we are asuming that the overall choice of duality frame between the parent and daughter theory is the same. If that is not the case and they are related by some $B\in Sp(2r, \mathbb{Z})$ then we would have $M_{\text{fixed}}=B V_NB^{-1}$} We prove this as follows. If the line $\mathcal{L}_*$ of the parent theory inherits the $\mathbb{Z}_2$ action (and hence has a fixed point), we can split its intersections with $\mathcal{S}$ into two sets of points: $u_{*,1},...,u_{*,n}$ and $-u_{*,1},...,-u_{*,n}$. The corresponding monodromies are, by \eqref{eq:electromagdualgeneral}, $M_{1},\dots,M_n$ and $V_N M_{1} V_N, \dots, V_N M_{n}V_N$. Consequently, using \eqref{eq:topologicalconditiongeneral}, we have 
\begin{align}
    \label{eq:topconditionparent} V_N M_1 \cancel{V_N V_N} M_2 V_N \cdots V_N M_n V_N M_{1} M_2 \cdots M_{n}  &= \left(V_N M_{1} M_2 \cdots M_{n}\right) \left(V_N M_{1} M_2 \cdots M_{n}\right) \nonumber\\
    &= M^2= M_{loop,parent}\,,
    \end{align}
     with $M \equiv V_N M_{1} M_2...M_{n}  $, with $M_{loop,parent}$ being the monodromy around a big anticlockwise loop inside $\mathcal L_*$ and we remark that we used that $V_N^2 = \mathbbm{1}$. By combining \eqref{eq:monodromyconditiongeneral} and  \eqref{eq:topconditionparent}, we must have
     \begin{equation}
     \label{eq:generalprooffromparent}
     M_{loop,daughter} = M = V_NM_{1} M_2\cdots M_{n}.
     \end{equation} We can now consider the topological condition \eqref{eq:topologicalconditiongeneral} for the monodromies in the daughter theory, obtaining 
    \begin{equation}
        \label{eq:topconditiondaughter}
        M_{\text{fixed}}M_1\cdots M_{n} =  M_{loop,daughter} 
    \end{equation}
    where, up to a global choice of electromagnetic duality frame, exactly the same monodromies $M_{j}$, $j = 1,\dots,N$ of the parent theory appear due to \eqref{eq:bpsspectragenres1}.  By inverting  \eqref{eq:topconditiondaughter}, and using \eqref{eq:generalprooffromparent}, we find 
    \begin{equation}
    \label{eq:z2monodromydef}
    M_{\text{fixed}} = V.
    \end{equation}
    Indeed, the preimage $u_{0}$ of $u_{\text{fixed}}$ under the quotient map is a smooth point of $\mathcal M_{\text{CB}}$, hence, in the parent theory, $M_{u_0} = \mathds{1}$. Consequently, using the argument of \cite{Argyres:2016yzz},  we must have $M_{\text{fixed}}^2 = \mathds{1}_r$. The interesting fact is that we do not need, and indeed we must not pick the trivial choice $M_{\text{fixed}} = \mathds{1}_r$.
    This predicts the existence of a singularity analogous to $(u_{\text{fixed}},M_{\text{fixed}})$ also for the $\mathbb Z_k$ gauging of a zero-form symmetry of a general (even non-lagrangian) 4d $\mathcal N = 2$ theory as explained in \cref{sec:generalresultzkspectra}.
    \item The last situation that may occur is when a fixed point of the parent theory is inside $\mathcal S$.

    We postpone the general analysis of the monodromies of these points to future work.  
\end{itemize}
In particular, \eqref{eq:bpsspectragenres1} permits to say that the generators of the BPS quiver computed from a generic slice of the Coulomb branch are preserved by the $\mathbb Z_2$ quotient.

\subsection{Explicit example: BPS spectrum of \texorpdfstring{$\widetilde{SU}(3)$}{}}
\label{sec:bpsspectrumsu3tilde}
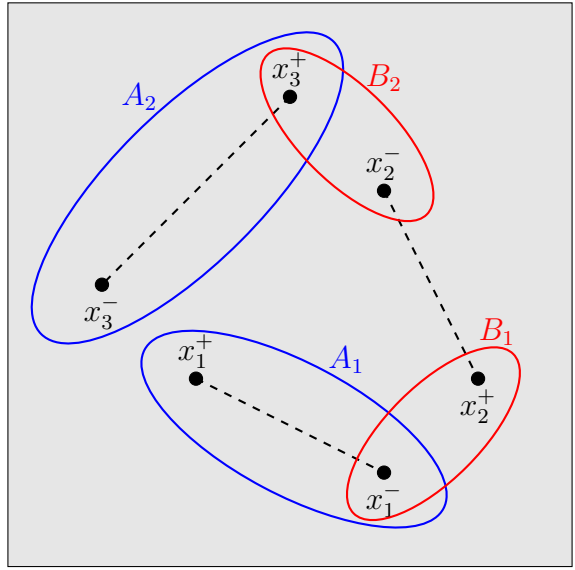
\begin{figure}[t]
\begin{center}
\begin{tikzpicture}[scale=2.5]
  % Draw plane
  \draw[fill=gray!20] (0,0) -- (3,0) -- (3,3) -- (0,3) -- cycle;

  % Draw points in pairs
  \foreach \x/\y in {1/1, 2/2, 1.5/2.5, 2.5/1, 0.5/1.5, 2/0.5} {
    \filldraw (\x,\y) circle (1pt);
  }

  % Group points
  \draw[thick,dashed] (1,1) -- (2,0.5);
  \draw[thick,dashed] (2,2) -- (2.5,1);
  \draw[thick,dashed] (1.5,2.5) -- (0.5,1.5);
  \node[above] at (1,1) {$x_1^+$};
  \node[above] at (2,2) {$x_2^-$};
  \node[above] at (1.5,2.5) {$x_3^+$};
  \node[below] at (2.5,1) {$x_2^+$};
  \node[below] at (0.5,1.5) {$x_3^-$};
  \node[below] at (2,0.5) {$x_1^-$};

%Cycles
  \draw[thick,blue,rotate=45] (2.1,0.75) ellipse (1.1 and 0.4);
\draw[thick,blue,rotate=-28] (1,1.36) ellipse (0.9 and 0.35);
\draw[thick,red,rotate=45](2.1,-1.1) ellipse (0.6 and 0.25);
\draw[thick,red,rotate=-45](-0.35,2.9) ellipse (0.6 and 0.25);

  \node at (0.7,2.5) {$\textcolor{blue}{A_2}$};
\node at (1.8,1.1) {$\textcolor{blue}{A_1}$};
    \node at (2.6,1.25) {$\textcolor{red}{B_1}$};
    \node at (2,2.6) {$\textcolor{red}{B_2}$};
\end{tikzpicture}
\end{center}
\caption{Choice of symplectic basis of 1-cycles, drawn in one of the two sheets of the SW curve. Note that the $B$ cycles (in red) only intersect the $A$ cycles (in blue) once, as they cross the branch cut and come back through the second sheet. Their orientations are such that $\langle A_i,B_i\rangle = +1$.}
\label{fig:sympbasiscyclesasinKlemm}
\end{figure}

In this section, we will compute explicitly the BPS spectrum of the theory $\widetilde{SU}(3)$. We recall, for convenience, the SW solution that we obtained in \cref{sec:SW_su3}:
\begin{equation}
\label{eq:swdatasu3tildasym}
y^2 = -\frac{\left(b^3+b u_2
u_6+u_6^2\right)^2}{u_6^3}+4 \Lambda ^6.
\end{equation}
The singular locus $\mathcal S_{SU(3)}$ of the parent $SU(3)$ SYM theory is 
\begin{eqnarray}
\label{eq:quantumdiscrsu3sym}
    \Delta_{SU(3)} &\propto&  216 u_2^3 \left(4 \Lambda ^6+u_3^2\right)+729
   \left(u_3^2-4 \Lambda ^6\right)^2+16 u_2^6 \nonumber \\
   &=& \left(108 \Lambda ^6-108
   \Lambda ^3 u_3+4 u_2^3+27 u_3^2\right)
   \left(108 \Lambda ^6+108 \Lambda ^3
   u_3+4 u_2^3+27 u_3^2\right). 
\end{eqnarray}
One can explicitly check, taking the discriminant of the r.h.s. of \eqref{eq:swdatasu3tildasym} w.r.t $b$, that the singular locus of the daughter theory is double-covered by the singular locus of the parent one being 
\begin{equation}
\label{eq:quantumdiscrsu3tildasym}
\Delta_{\widetilde{SU}(3)}  \propto  216 u_2^3 \left(4 \Lambda ^6+u_6\right)+729
   \left(u_6-4 \Lambda ^6\right)^2+16 u_2^6 = \Delta_{SU(3)}(u_3 \to \sqrt{u_6})\,.
\end{equation}
We can observe the following facts by matching the two discriminant loci:
\begin{itemize}
    \item The double-cover map, as expected, is the one that sends $u_2 \to u_2$ and $u_3 \to u_6 = u_3^2$, and is induced by the $\mathbb Z_2$ action on the parent theory Coulomb branch;
    \item $\Delta_{SU(3)}$ displays two irreducible components, \begin{equation}
    \label{eq:irredcompsu3discrlocus}
    f_1 = 108 \Lambda ^6-108 \Lambda ^3 u_3+4
   u_2^3+27 u_3^2, \qquad f_2  = f_1 \rvert_{u_3 \to -u_3}.
   \end{equation}
   These two components are identified by the $\mathbb Z_2$ discrete gauging action into the single irreducible component of the r.h.s. of \eqref{eq:quantumdiscrsu3tildasym}
   \end{itemize} 
   Finally, we fix the same symplectic basis of one-cycles as in \cite{Klemm:1994qj}, see Figure \ref{fig:sympbasiscyclesasinKlemm}. 

We are now ready to check the expectations of \cref{sec:generalbpsspectra} with an explicit computation. 
It is important to notice that the slices $u_3 = \text{const.}$ and $u_6 = \text{const.}$ are generic in their respective Coulomb branches and intersect the respective bifurcation loci in six points. On the opposite, the slice $u_2 = \text{const.}$ are non-generic in both the $SU(3)$ and $\widetilde{SU}(3)$ Coulomb branches, as they intersect the bifurcation locus, respectively, just in four and two points. \\

Let us start recalling the results of \cite{Klemm:1994qj}. We concentrate on the $u_2 = \text{const}$ slice in the $SU(3)$ Coulomb branch, where the coordinate $u_3$ is free to vary and describes a complex plane that intersects $\mathcal S$ in six points, that we call $u_{3,1},...,u_{3,6}$.
The charges of the BPS particle becoming massless at a certain $u_{3,k}$ can be computed using the procedure outlined before \eqref{eq:topologicalconditiongeneral}.

For the considered line, we obtain\footnote{Note that the overall sign of the charges of a BPS particle vanishing at a point is arbitrary: the monodromy associated with the cycle $\nu$ is the same of the one associated with $-\nu$.}
\begin{eqnarray}
\label{eq:bpsparticlessu3u2plane}
    q_{3,1} = (1,0,1,0), \quad q_{3,2} = (-1,-1,0,-1), \quad q_{3,3} = (0,1,-1,1), \nonumber \\
        q_{3,4} = (-1,0,1,-1), \quad q_{3,5} = (0,-1,0,1), \quad q_{3,6} = (1,1,-1,0).
\end{eqnarray}
The points particles $u_{3,1},u_{3,2},u_{3,3}$ are related by the $\mathbb Z_3$ action $u_2 \to e^{\frac{2\pi i}{3}}u_3$, and consequently \cite{Klemm:1994qj,Klemm:1995wp}, their charges are related as
\begin{eqnarray}
\label{eq:electromagntransfsu3z3}    q_{3,1} &=& q_{3,2}U,\quad q_{3,2} = q_{3,3}U,\quad q_{3,3} =  q_{3,1}U,  \nonumber \\
    q_{3,4} &=& q_{3,6}U,\quad q_{3,5} = q_{3,4}U,\quad q_{3,6} = q_{3,5}U,
\end{eqnarray}
with $U$ being the following electro-magnetic duality transformation\footnote{Our conventions match those of \cite{Klemm:1994qj}, that differ from the ones of \cite{Klemm:1995wp} where this effect is explicitly mentioned. The convention of \cite{Klemm:1995wp} can be matched with our convention using $U = \eta^{-1} U_{\text{there}}\eta$, with 
\begin{equation*}
   \eta = \left(
\begin{array}{cccc}
 0 & 1 & 0 & -1 \\
 1 & 0 & 1 & 0 \\
 0 & 0 & 0 & -1 \\
 0 & 0 & -1 & 0 \\
\end{array}
\right)
\end{equation*}
}
\begin{equation}
    \label{eq:electromagdualz3}
U = \left(
\begin{array}{cccc}
 0 & 1 & 0 & 0 \\
 -1 & -1 & 0 & 0 \\
 0 & 0 & -1 & 1 \\
 0 & 0 & -1 & 0 \\
\end{array}
\right)
\end{equation}
One can explicitly check that $q_{3,2}$ and $q_{3,4}$ can be written as combinations of the other charge vectors. Computing the intersection pairing between $q_{3,1},q_{3,3},q_{3,5},q_{3,6}$  one finds
\begin{equation}
\label{eq:diracpairingsu3}
\left(
\begin{array}{cccc}
 0 & -1 & 0 & -2 \\
 1 & 0 & 2 & 0 \\
 0 & -2 & 0 & -1 \\
 2 & 0 & 1 & 0 \\
\end{array}
\right)\,.
\end{equation}

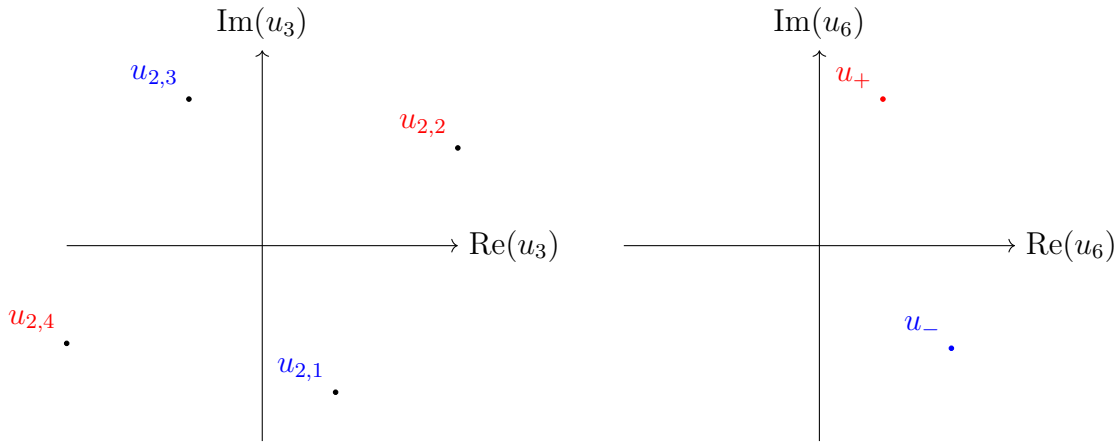
\begin{figure}
\centering
\begin{subfigure}[t]{0.4\textwidth}
\begin{tikzpicture}[scale=0.65]
  % Axes
  \draw[->] (-4,0) -- (4,0) node[right] {$\mathrm{Re}(u_3)$};
  \draw[->] (0,-4) -- (0,4) node[above] {$\mathrm{Im}(u_3)$};

  % Marked points
  \foreach \point/\label in {(1.5,-3)/u_{2,1}, (-1.5,3)/u_{2,3}} {
    \filldraw \point circle (1.3pt);
    \node[anchor=south east, blue] at \point {$\label$};
  }
   \foreach \point/\label in {(4,2)/u_{2,2}, (-4,-2)/u_{2,4}} {
    \filldraw \point circle (1.3pt);
    \node[anchor=south east, red] at \point {$\label$};
  }
\end{tikzpicture}
\end{subfigure}
\qquad \qquad
\begin{subfigure}[t]{0.4\textwidth}
 \begin{tikzpicture}[scale=0.65]
  % Axes
  \draw[->] (-4,0) -- (4,0) node[right] {$\mathrm{Re}(u_6)$};
  \draw[->] (0,-4) -- (0,4) node[above] {$\mathrm{Im}(u_6)$};

  % Marked points
  \foreach \point/\label in {(1.3,3)/u_{+}} {
    \filldraw[red] \point circle (1.3pt);
    \node[anchor=south east, red] at \point {$\label$};
  }
   \foreach \point/\label in {(2.7,-2.1)/u_{-}} {
    \filldraw[blue] \point circle (1.3pt);
    \node[anchor=south east, blue] at \point {$\label$};
  }
\end{tikzpicture}   
\end{subfigure}
\caption{Singular points on the $u_2 =$ const. slice of the $SU(3)$ and $\widetilde{SU}(3)$ Coulomb branch, respectively. We color code the $\mathbb{Z}_2$ orbits in the parent theory and to which point they descend after the discrete gauging.}
\label{fig:u3planesu3}
\end{figure}

Let us then pass to the $u_3 = \text{const.}$ slice in the SU(3) Coulomb branch. On this slice, the coordinate $u_2$ is free to vary and describes a complex line that intersects $\mathcal S$ in four points that we depict in figure \cref{fig:u3planesu3},
that we call $u_{2,1}, u_{2,2},u_{2,3}, u_{2,4}$. One can check that, for our choice of the basis of one-cycles, the charges $q_{2,k}$ of the BPS particles becoming massless at $u_{2,k}$ are
\begin{equation}
    \label{eq:bpsparticlessu3u3plane}
    q_{2,1} = (0,-1,0,1), \quad q_{2,2} = (-1,-1,1,0), \quad q_{2,3} = (-1,0,-1,0), \quad q_{2,4} = (0,-1,1,-1), 
\end{equation}
where we ordered the points as in \cref{fig:u3planesu3}.
We remark that the number of intersection points $u_{2,k}$ is less than the one of the $u_3 = \text{const}$ slice, hence the $u_2 = \text{const}$ slice is non-generic. However, one can check that \eqref{eq:bpsparticlessu3u2plane} are linearly independent over $\mathbb Z$ and the Smith normal form of their Dirac pairing is the same of \eqref{eq:diracpairingsu3}.
As pointed out in \cite{Klemm:1995wp}, the four singularities can be divided into two pairs, each pair being a $\mathbb Z_2$ orbit. Then, as predicted by \eqref{eq:electromagdualgeneral}, the electromagnetic charges of the BPS becoming massless are related by the right action $q_{2,1} = q_{2,3}V$ and $q_{2,2} = q_{2,4}V$, where
   \begin{equation}
\label{eq:z2electromagneticduality}
   V \equiv \left(
\begin{array}{cccc}
 0 & 0 & 0 & -1 \\
 0 & 0 & 1 & 0 \\
 0 & 1 & 0 & 0 \\
 -1 & 0 & 0 & 0 \\
\end{array}
\right)\,.
   \end{equation}

Let us now consider the daughter $\widetilde{SU}(3)$ theory, we can start from the slice at $u_6 = \text{const}$. By explicit computation, one can check that there are six points $\widetilde{u}_{2,k}$ where the slice intersects $\mathcal S$. The preimages of $\widetilde{u}_{2,k}$ inside $\mathcal M_{\text{CB}}$ are not fixed points of the $\mathbb Z_2$ quotient action\footnote{We are considering the slice $u_{6}=c_0$, this corresponds to two different slices $u_3 = \pm \sqrt{c_0}$ with an empty intersection that get identified by the $\mathbb Z_2$ action.} and that, as predicted in \cref{sec:generalbpsspectra}, the charges of the massless BPS particles at $\widetilde{u}_{2,k}$ are exactly \eqref{eq:bpsparticlessu3u2plane}. We can again isolate the generators, and compute the BPS quiver intersection matrix by computing their Dirac paring, obtaining \eqref{eq:diracpairingsu3}. Consequently, we just showed that the BPS quiver is preserved by the discrete-gauging procedure.

To conclude, we consider the more interesting slice at $u_2 = \text{const}$, (with $u_6$ free to vary). There are just two points, $u_+$ and $u_-$, depicted in \cref{fig:u3planesu3} where we have a massless BPS particle, and the two charges are 
\begin{equation}
\label{eq:bpsparticlessu3tildau2plane}
    q_{-} = (-1,-1,1,0), \qquad q_{+} = (0,-1,0,1).
\end{equation}
Apart from these ``classic'' BPS particle degeneration, the $u_{2} = \text{const.}$ slice displays a peculiar feature: it intersects the locus $u_6 = 0$ where the SW geometry \eqref{eq:swdatasu3tildasym} develops a pole in the Coulomb branch coordinates. 
This point is peculiar also because its preimage inside $\mathcal M$ is a fixed point of the $\mathbb Z_2$-quotient map. This is due to the fact that, contrary to the $u_3 = \text{const}$ slice, the $u_2 = \text{const}$ slice is an $\mathbb Z_2$ invariant subset of the parent theory Coulomb branch, and hence, after the quotient, possesses a fixed point at $u_{6} = 0, u_2 = \text{const}$. We can again compute its monodromy by looking graphically at how the branching points of the hyperelliptic double cover get permuted going anticlockwise around $u_6 = 0$ obtaining
\begin{equation}
    \label{eq:devilequation}
    M_{\text{fixed}}= V.
\end{equation}
Stated differently, the result is the following: \textit{a non-singular fixed point of the $\mathbb Z_2$ action of the parent theory gets mapped to a singular point $p_{\text{fixed}}$ of the daughter theory Coulomb branch. The monodromy  $M_{\text{fixed}}$ around $p_{\text{fixed}}$ is exactly the electromagnetic duality $V$ realizing the $\mathbb Z_2$ action in the parent theory.}\\ 
\indent On top of the general proof in \cref{sec:generalbpsspectra}, we can understand heuristically the situation as follows. We should expect something peculiar to happen on the $\widetilde{\mathcal L_{u_6}} \equiv \left\{u_{2} = \text{const}\right\}$ slice: we can consider the restriction of Riemann-Hilbert problem associated with the SW solution to  $\widetilde{\mathcal L_{u_6}}$. The slice is the $\mathbb Z_2$ quotient of the $\mathcal L_{u3} \equiv u_2 = \text{const}$ slice of the parent theory. By passing from $\mathcal L_{u_3}$ to $\widetilde{\mathcal L_{u_6}} = \mathcal L_{u_3}/\mathbb Z_2$  we halved the number of points in which we have a BPS massless particle (as we identified them pairwise during the quotient). Hence, the Riemann-Hilbert problem restricted to the considered slice is ill-defined, unless we introduce something else to balance the fact that we halved the number of points supporting a monodromy and that we changed the monodromy at infinity as in \eqref{eq:monodromyconditiongeneral}. Indeed, heuristically this boils down to the fact that, having two choices in \eqref{eq:bpsspectragenres1}, we have multiple equivalent ways, depending on the choices branch cuts of $x_k(u,m)$, to relate the monodromies of the daughter theory to those of the parent one. In some sense, $V_{\text{fixed}}$ permits us to move between these equivalent choices.

\subsection{BPS spectra for \texorpdfstring{$\mathbb Z_k$}{} gauging of general 4d $\mathcal{N}=2$ theories}
\label{sec:generalresultzkspectra}
In this section, we generalize the presence of an object analogous to the pair $(u_{fixed},M_{fixed})$ that we introduced around \eqref{eq:z2monodromydef} for 
the $\mathbb Z_k$ discrete-gauging, with $k>2$, of a generic $\mathcal N =2, \mathcal D = 4$ theory. The argument goes in the same way as in the $\mathbb Z_2$ lagrangian case: in the parent theory, we have an electromagnetic duality $\mathcal E$ generalizing the action of $V$, with 
\begin{equation}
\label{eq:generalelectromagneticdual}
\mathcal E^k = \mathds{1}_{2r}
\end{equation}
with $r$ the rank of the theory. Notice that \eqref{eq:generalelectromagneticdual} is a simplyfing assumption, as in general we could have e.g. $\mathcal E^{k}= -\mathds{1}$.

\begin{figure}\begin{center}
\includegraphics[width=8cm]{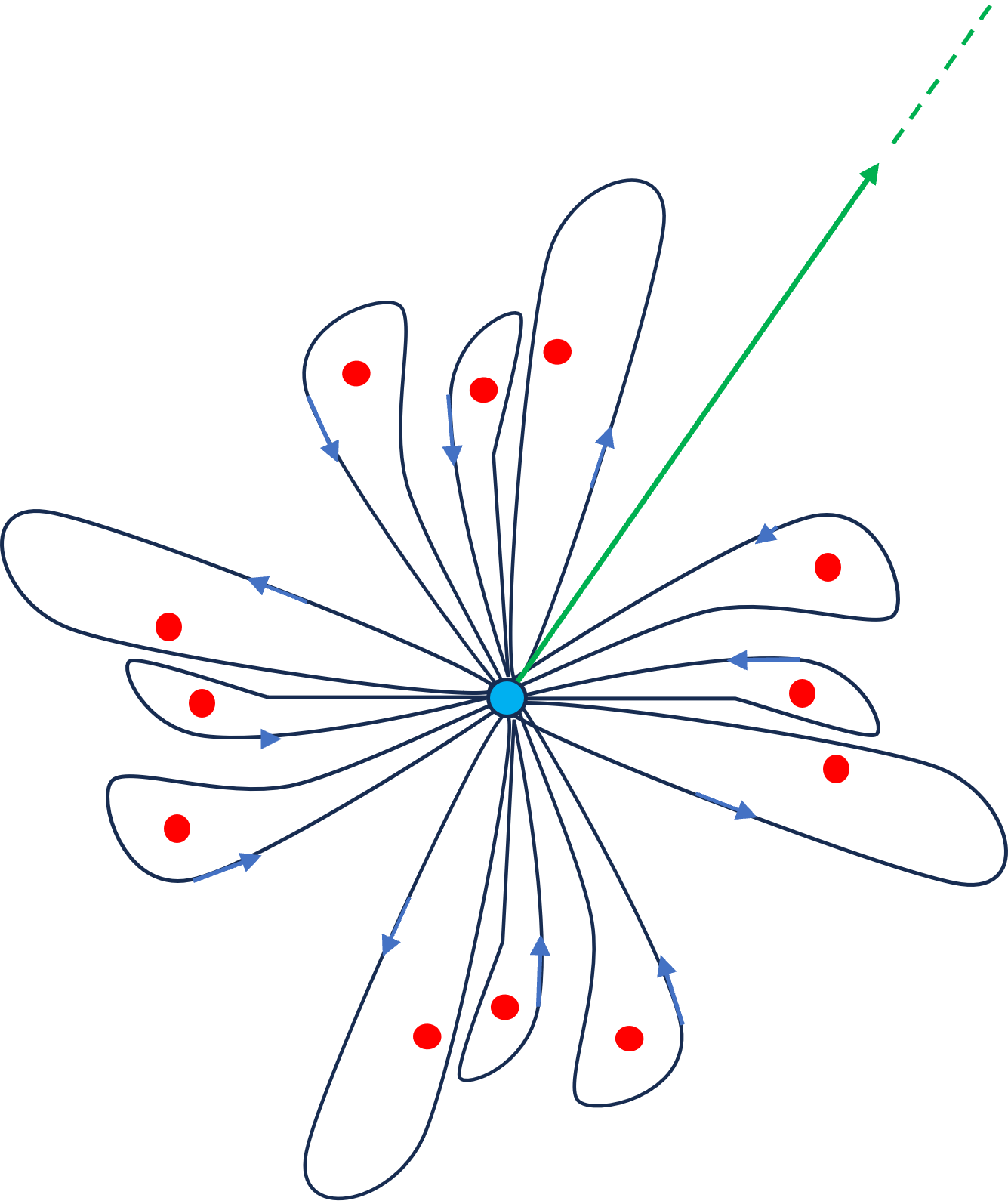}
\caption{Clustering of singular points in a Coulomb branch slice $\mathcal L$ of a theory enjoying a $\mathbb Z_k$ symmetry (here $k = 4$). The singular points are indicated in red, and the first "cluster" is formed by the three rightmost red points. The light-blue point corresponds to a base-point for an anticlockwise loop obtained composing the tiny blue loops. Hence, the monodromy on the big loop is the product of the monodromies associated with the red points. The monodromies are ordered from the right to the left according to the sequence in which the red points get hit when we rotate anticlockwise the green ray emanating from the light-blue point.}
\label{fig:bigloop1234}
\end{center}

\end{figure} 

Let's now consider a complex line $\mathcal L$ inside  $\mathcal M_{\text{CB}}$, such that $\Gamma\rvert_{\mathcal L} \in \text{Aut}(\mathcal L)$. Then, calling $U$ the coordinate on $\mathcal L$, being the $\Gamma$ action linear, we have just one fixed point at $U = 0$. The points $U_{i} \in \mathcal L$ where we have massless BPS particles organize, as in Figure \ref{fig:bigloop1234}, into $k$ clusters (related by the action of $\Gamma$), that become more and more separated as we take large values for the Coulomb branch parameters that we fixed\footnote{E.g., in the $SU(3)$ example of \cref{sec:bpsspectrumsu3tilde}, we are taking $\abs{u_2} >> 1$ for the $u_2 = \text{const.}$ slice of the parent theory Coulomb branch.} in order to obtain $\mathcal L \subset \mathcal M_{\text{CB}}$. Let's call $U_{t}$, with $t= 1,...n_{cluster}$ the points belonging to the first cluster, and consider a point $\Gamma^j(U_{t})$ belonging to the $j$-th cluster. $\Gamma^j(U_{t})$ is obtained from $U_t$ applying $j$ times the map $\Gamma$, then 
\begin{equation}
    \label{eq:chargesrelationsgeneral}
    q_{\Gamma^j(U_{t})} =  q_{U_{t}} \mathcal E^j, \quad M_{\Gamma^j(U_{t})} = \mathcal E^{-j} M_{U_{t}} \mathcal E^j. 
\end{equation}
The monodromy on a big anticlockwise loop on $\mathcal L$ is 
\begin{equation}
    \label{eq:bigloopgeneral}
     M_{loop,parent} = \mathcal E^{-k+1}M_{cluster} \mathcal E^{k-1}\mathcal E^{-k+2} \cdots M_{cluster}, 
\end{equation}
with $M_{cluster} = \prod_{t=1}^{n_{cluster}}M_{U_{t}}$. By using  $\mathcal E^k = \mathds{1}$, \eqref{eq:bigloopgeneral} can be rewritten as 
\begin{equation}
    \label{eq:smartmodification}
    M_{loop,parent} = (\mathcal E M_{cluster})^k.
\end{equation}
The monodromy of the daughter theory on a big anticlockwise loop is related to the one of the parent theory by 
\begin{equation}
\label{eq:monodromyrelgeneral}
M_{loop,daughter}^k = M_{loop,parent}
\end{equation}
The only way in which \eqref{eq:smartmodification} and \eqref{eq:monodromyrelgeneral} are consistent is that 
\begin{equation}
    \label{eq:finalsol}
    M_{loop,daughter} = \mathcal E M_{cluster},
\end{equation}
where we used that the ``roots of units'' ambiguity given by taking the $k-$th. root is not present because we are looking for integer-valued solutions\footnote{One might be worried when some of the roots of unit are real, in that case they must be $-1$, and this simply encodes the fact that, on a given $U_{t}$, both $\nu $ and $-\nu$ vanishes at the same time. In other words, in that case $\mathcal E$ can be equally re-defined as $- \mathcal E$.}.
Now, by using that again, for $U\neq 0$ the $\mathbb Z_k$ action $\widetilde{\Gamma}$ identifies different curves (rather than taking a quotient of the same fiber of the parent theory SW solution) of the parent theory SW family, we have $\widetilde{M_{t}} = M_{t}$ at points $\widetilde{U}_t \in \widetilde{\mathcal L}$. Consequently, we get
\begin{equation}
\label{eq:loopdaughterfinalres}
M_{loop,daughter}  = M_{fixed}\prod_{t=1}^{n_{cluster}} M_t = M_{fixed} M_{cluster}
\end{equation}
and hence, taking \eqref{eq:finalsol} and \eqref{eq:loopdaughterfinalres}  together, we have 
\begin{equation}
    \label{eq:finalres}
    M_{fixed} = \mathcal E.
\end{equation}
We can reformulate the result as follows: \\
\indent Let $\mathcal L$ be a slice of the parent theory Coulomb branch that inherits the $\mathbb Z_k$ discrete gauging action. Let $U$ being its coordinate, and $\widetilde{U}$ the coordinate on $\widetilde{\mathcal L} = \mathcal L/\mathbb Z_k$. Finally, let $\mathcal E$ be the electromagnetic duality realizing the $\mathbb Z_k$ action at the level of the fibers of the SW geometry of the parent theory. Then, if $\mathcal E^k = \mathds{1}$, we get a monodromy equals to $\mathcal E$ at $\widetilde{U} = 0$.

To conclude, we note that this result is again compatible with the argument of \cite{Argyres:2016yzz}: the preimage under the $\mathbb Z_k$ quotient of $u_{fixed} \in \widetilde{\mathcal M_{\text{CB}}}$ is, in our framework, a smooth point $u_0 \in \mathcal M_{\text{CB}}$ fixed by the $\mathbb Z_k$ action. Consequently, going once around $u_{0}$ corresponds to go $k$ times around $u_{\text{fixed}}$, hence
\begin{equation}
    \label{eq:compatibilitycheckzk}
    M_{\text{fixed}}^k = M_{u_0} = \mathds{1}_r,
\end{equation}
where in the last passage we used that $u_0$ is a smooth point of $\mathcal M_{\text{CB}}$. Our argument shows that, if $\mathcal E^k = \mathds{1}_r$, the monodromy of $u_{fixed}$ is non-trivial and corresponds to $\mathcal E$.

\section{Discussion and outlook}\label{sec:conclusion}

Let us summarize our results. We have found the SW solutions for theories with disconnected gauge groups $\widetilde{SU}(N)$ starting from those of the theories with $SU(N)$ gauge group. Our method consists of finding the lift of the (known) action $\Gamma$ on the Coulomb branch to the full SW geometry, by requiring that $\widetilde{\Gamma}$ 1) commutes, in the conformal case, with the $U(1)_R$ action, 2) is an automorphism of the SW geometry, and 3) preserves the SW differential. Then, one can express the SW solution before the discrete gauging in terms of the $\mathbb Z_2$ invariant coordinates and impose the new relations that come from the quotient.

From these solutions, we also studied the singularities on the Coulomb branch where extra particles become massless. Most of these are inherited in a straightforward way from the theory before the discrete gauging; however, our SW solutions show the existence of ``new'' points where the naive EFT fails that descend from smooth $\Gamma$-invariant points of Coulomb branch of the theory with gauge group $SU(N)$. We also computed the monodromy $M_{\text{fixed}}$ around these ``new'' points to be exactly the electromagnetic duality $V_N$ realizing $\widetilde{\Gamma}$ in the theory before the discrete gauging. We then rephrased the appearance of these new singularities in terms of the interplay between the Riemann-Hilbert problem and the discrete-gauging procedure, suggesting them to be present also for the discrete-gauging of non-lagrangian theories. 

Let's now proceed to discuss, as mentioned in the introduction, some implications of our results.

\subsection{On the classification of rank-2 geometries}

As we discussed in detail in the main text, one of the consequences of the quotient at the level of the SW geometry is that it can lead to poles on the Seiberg-Witten curve. For convenience, we reproduce here the curve for $\widetilde{SU}(3)+6 (F\oplus\overline{F})$,
\begin{align}\label{eq:recapsu3}
    y^2 &= -4f a^3 +a^3+2 u_2 \left(a^2+b\right)+2 a
   b+a u_2^2+u_6\,,\nonumber\\
   b^2 &= a u_6\,.
\end{align}
The presence of $1/u_6$ terms when substituting $a$ in terms of $b$ into the first equation is what explains why this theory, with Coulomb branch parameters of dimensions $(2,6)$ and rank 3 global symmetry, did not appear in the classification of Argyres and Martone \cite{Argyres:2022lah}.

It is interesting to remark that the new singularity at $u_6=0$ with $u_2=const$, produced by the discrete gauging, turns out to not be very exotic. Indeed, we can plug $u_6=0$ into \eqref{eq:recapsu3}, which results in $b=0$ and $a$ being a free coordinate. Then, the first equation becomes
\begin{align}
    y^2 = (1-4f) a^3 + 2u_2 a^2 + a u_2^2\,,
\end{align}
which is a very simple elliptic curve. Since the genus has dropped to 1, and the only CB parameter has scaling dimension 2, we can identify this as the $SU(2)+4 F$ theory.

More generally, our computation shows that it is quite easy to find theories where the SW geometry shows coefficients which are non-polynomial in the Coulomb branch parameters, and as a consequence, extending the classification of \cite{Argyres:2022lah} to geometries where poles are allowed is an interesting open problem.

\subsection{Comments on SymTFT for non-invertible defects}\label{sec:comments}

One of the interesting features of theories with disconnected gauge groups is that, if the matter fields live only in the adjoint representation, they have a non-invertible 1-form symmetry \cite{Heidenreich:2021xpr,Arias-Tamargo:2022nlf,Bhardwaj:2022scy}. It is therefore interesting to explore whether or not it can be seen from the SW solution. In \cite{DelZotto:2022ras}, a procedure was found to extract the SymTFT for the (grouplike) 1-form symmetry from the SW geometry. One starts from a BF theory in 5d that schematically looks like
\begin{align}
    S_{\text{SymTFT}} = 2\pi i\, N \int_{M^5} B\wedge dC\,,
\end{align}
where $B$ and $C$ are 2-form fields that serve as a background connection for the electric and magnetic 1-form symmetry. Then, the idea is to find which are the local particles that can screen the relevant line operators. In the end, this results in a very simple recipie: the coefficients that we need to fix the SymTFT can be extracted from the BPS quiver by computing the Smith Normal Form of the intersection pairing between the vanishing cycles. In the example of $SU(3)$, the intersection matrix is \eqref{eq:diracpairingsu3}, which has Smith Normal Form
\begin{align}
    \left(
\begin{array}{cccc}
 1 & 0 & 0 & 0 \\
 0 & 1 & 0 & 0 \\
 0 & 0 & 3 & 0 \\
 0 & 0 & 0 & 3 \\
\end{array}
\right)\,.
\end{align}
This is telling us that the relative theory has a $\mathbb{Z}_3\times \mathbb{Z}_3$ 1-form symmetry and upon a choice of polarization there remains a $\mathbb{Z}_3$ either electric or magnetic 1-form symmetry; this is encoded on the SymTFT
\begin{align}
    S_{\text{SymTFT}} = 6\pi i \int_{M^5} B\wedge dC\,.
\end{align}

Now, as we have discussed in Section \ref{sec:bps_quiver}, the intersection matrix between the vanishing cycles is preserved when performing the quotient. This seems to lead us to the conclusion that the SymTFT should be the same; however, we know that this cannot be all as the 1-form symmetry of $\widetilde{SU}(3)$ is non-invertible. Indeed, one should note that the analysis of the screening of line operators is the same for the connected and disconnected gauge groups, and what gives rise to the non-invertibility is the way one builds gauge invariant topological defects after the discrete gauging. Therefore, it is not a contradiction that the procedure of \cite{DelZotto:2022ras} doesn't detect this. Investigating what is the way to generalize that procedure in such a way as to be able to detect the non-invertibility from the SW curve is a very interesting open problem. It is tempting to guess that the new singularity introduced by the quotient, which precisely accounts for the ambiguity of choice of pre-image of a point in the parent theory Coulomb branch, should play an important role in this. 

\subsection{String and M-theory uplift}\label{sec:M_theory_uplift}

It is well known that there is an intimate relation between the Seiberg-Witten solution of a theory and it's engineering in String Theory \cite{Witten:1997sc}. Therefore, one can try to use this relation to engineer the theories with disconnected gauge groups.\footnote{In this direction, it should be noted that the case of $O$ groups (which are built from gauging the outer automorphism of a $D$-type algebra) is known and $\widetilde{SU}(N)$ seems to be a more complicated problem \cite{GarciaEtxebarria:2022vzq,Bergman:2022otk}.}

The pure $SU(N)$  gauge theory can be realized in type IIA as the world-volume theory of $N$ D4 suspended between two vertical NS5 branes. Both the D4s and the NS5s get uplifted in M-theory to M5 branes, that recombine into a single M5 brane with a non-trivial worldvolume \cite{Witten:1997sc}. In particular, the recombined M5 brane expands to infinity at the points corresponding to the NS5 locations. In this setup, the SW curve \eqref{eq:swcurvesunfflavorsgeneral} describes the worldvolume of the recombined M5 along the $\mathbb C^2 \ni (x,y)$ (one, in general, has to add points at infinity, obtaining the class-S construction \cite{Gaiotto:2009we}).

In this setup, the reflection of the coordinates $(x,y)$ has a physical meaning. We can start with the $N$-odd case, in which just the coordinate $x$ gets reflected. We set, for simplicity, the masses to zero, but the arguments proceed in the same way with the caveat of turning on only the mass-deformations that respect the $\mathbb Z_2$-symmetry.  The world-volume that corresponds to the Coulomb branch of the daughter theory is then \eqref{eq:swcurvesunfflavorsgeneral}, with the points $(x,y)$ and $(-x,y)$ identified. In this case, the reflection of a single coordinates transforms $\mathbb C^2 \ni(x,y)$ into $\mathbb C^2 \ni(x^2,y)$, that is still a Calabi-Yau surface, and then it makes sense for the setup to be still $\mathcal N = 2$. The situation is analogous for the $N$-even case, with the caveat that, since now the $\mathbb Z_2$ action reflects both $x$ and $y$, the M5 brane is contained in $\mathbb C^2 /\mathbb Z_2 \cong A_1$, the rank-one Du Val singularity\footnote{In our opinion, this setup is still different from the one presented in \cite{Witten:1997sc}, where an $A_1$ singularity corresponded to the presence of D6 branes in the IIA limit. Indeed, in \cite{Witten:1997sc} the metric on the $A_1$ singularity is the Taub-NUT one, also called ALF metric. In our case, instead, the most natural metric is the ALE one, that asymptotically reproduces the flat orbifold $\mathbb C_2/\mathbb Z_2$.}.

Apart from the IIA/M-theory construction with branes, the SW solution is also related to type IIB geometric engineering \cite{Katz:1997eq,Tachikawa:2011yr}. We can consider IIB on  the following threefold: 
\begin{equation}
\label{eq:threefoldeng}
a b = \Sigma(x,y)
\end{equation}
with $\Sigma(x,y)$ the SW curve of the theory, and $(a,b,x,y) \in \mathbb C^4$, where the role of the SW differential is played by the holomorphic volume form of \eqref{eq:threefoldeng}. In this case, again, we see that the $\mathbb{Z}_2$ action at the level of the SW curve becomes a reflection of spacetime coordinates in the type IIB embedding.

The problem however, is that the reflection of $x,y$ for us is accompanied with a reflection on the $u_j$. In our construction, that is at the geometric level of the SW solution, we are treating the variables $u_j$ and $(x,y)$ on an equal footing (e.g., by constructing invariants that mixes $u_j$ and $(x,y)$). In the string theory uplifts though, the $u_j$ and the $(x,y)$ are completely different in nature, being the first parameters of the physical system, while the second are space-time coordinates. One is reminded of a similarity with S-folds in type IIB, where a spacetime action is combined with a non-trivial action on the axio-dilaton $\tau$. However, in that context, F-theory precisely geometrizes $\tau$ in such a way as to be able to treat it as well as the coordinates on equal footing. At this stage, we leave a deeper investigation of the M-theory realization of disconnected gauge theories for future work.

\subsection{Open problems}

There are more open questions that our investigation opens the door to, beyond the ones discussed above. Let us comment on a few of them:
\begin{itemize}

\item One interesting point would be to delve into the dynamics of the theory from of our SW solution, namely, use them to try to compute the prepotential. This is a natural next step, as so far our investigation of the dynamics of the disconnected gauge theory was just scratching the surface by looking at the particles that become massless. Since, in the end, the SW technology is a smart way to compute the instanton contributions to the partition function, it would be interesting to try to use this to understand in greater generality the instantons of disconnected gauge groups.

\item Related to the previous point, we have seen that the appearance of new singularities descending from \textit{smooth} points of the theory before the discrete gauging theory fixed by the quotient action should be a very general feature of any discrete gauging. Understanding the physics of these Coulomb branch points, which at the moment we understand just as a natural requirement coming from the Riemann-Hilbert picture, would be a tantalizing follow-up of this work. Moreover, as we have discussed, we expect this effect to take place also for discrete gaugings of non-lagrangian theories. Another direction worthy of exploration would be to study the singularities that arise in this way in more exotic examples. As we discussed above, for $\widetilde{SU}(N)$ the theory living on that singularity was simply $SU(2)+4 F $, but more interesting theories may be found in general.

\item We believe that the phenomena of complex structure singularities in the CB is deserving of further study beyond our simple example. Typically in mathematical literature \cite{Alastair:2002yt}, a possible way to resolve the complex structure singularities of a given moduli space $\mathcal M$ parametrizing a family of objects $\mathcal F$ is to find a smart, and in some sense ``minimal'', rephrasing of the definition of the objects in $\mathcal F$. For example, we can construct the resolution of the $A_1 \cong \mathbb C^2/\mathbb Z_2$ singularity as follows:  consider the symmetric product of $\mathbb C^2$ with itself; the $\mathbb Z_2$ action on $\mathbb C^2$ induces a natural $\mathbb Z_2$ action on the symmetric product. One of the invariant components inside the symmetric product w.r.t the induced $\mathbb Z_2$ action is the $A_1$ singularity. Now, to produce the resolution we perform the same procedure, but using the ``Hilbert scheme of two points in $\mathbb C^2$'', that contains the symmetric product of $\mathbb C^2$ with itself as a dense open subset. After considering again the $\mathbb Z_2$ invariant part of the Hilbert scheme, we get now the \textit{resolved} $A_1$ singularity. Stated differently, resolving the $A_1$ singularity is equivalent, in this example, to modifying the notion of objects parametrized by it, passing from varieties (in this case, couples of distinct antipodal points) to schemes. It would be interesting to understand whether this logic applies to our situation, and what would be the relevant ``object'' such that the complex structure singularity of our examples becomes resolved.

\item Lastly, throughout the paper we have selected one of the two possible ways to gauge charge conjugation in a $SU(2N)$ theory \cite{Arias-Tamargo:2019jyh}. Understanding the SW solution for the other possible discrete gauging, as well as the relation, if there is one, with the apparent difference between the action on the coordinates $x,y$, is another open problem.
    
\end{itemize}

%---------- Acknowledgements ------

\subsection*{Acknowledgments}

It's a pleasure to thank Michele Del Zotto, Pietro Longhi, Kunal Gupta,  Robert Moscrop, Guglielmo Lockhart, Iñaki García-Etxebarria, Sergej Monavari, Federico Carta and Amihay Hanany for discussions. We particularly thank Roberto Valandro for comments on the draft. GAT is supported by the STFC Consolidated Grants ST/T000791/1 and ST/X000575/1. 
The work of MDM is supported by the European Research Council (ERC) under the European Union’s Horizon 2020 research and innovation program (grant agreement No. 851931).
MDM thanks the ``Fondazione Angelo Della Riccia'' for financial support during the initial stage of this work. 
 
\vspace{1cm}
%\clearpage
%--------- Appendix --------------

\begin{appendix}
\section{ The case of \texorpdfstring{$SU(2)+ 4\, F$}{}} \label{sec:SW_su2}
In this appendix, we will apply the IR geometric procedure of the main text to the case of $SU(2) + 4F$, reproducing the result of \cite{Argyres:2016yzz}. This will serve as a non-trivial test for our method, and will show it at work in a different setup, where the discrete symmetry arises promoting an S-duality of the UV theory to a symmetry by finely tuning the UV coupling.

The SW data before the discrete gauging are \cite{Seiberg:1994aj}: 
\begin{equation}
\label{eq:startingswdatasu2}
    y^2 = (x-e_1 u)(x-e_2 u)(x+(e_1+e_2) u), \quad \lambda = u \frac{dx}{y},  
\end{equation}
with $e_i$ being defined by
\begin{equation}
    \label{eq:thetafunctions}
    e_1-e_2 = \left(\theta_3(0,\tau)\right)^4, \quad e_3-e_2 = \left(\theta_1(0,\tau)\right)^4, \quad e_1-e_3 = \left(\theta_2(0,\tau)\right)^4,
\end{equation}
where $\theta_j$ are the Jacobi $\theta$-functions on the elliptic curve of modular parameter $\tau$,
\begin{eqnarray}
    \label{eq:thetafunctions2}
    \theta_1(0,\tau) = \sum_{n\in \mathbb Z} q^{1/2(n+1/2)^2},   \quad  \theta_2(0,\tau) = \sum_{n\in \mathbb Z} (-1)^n q^{1/2n^2},     \quad \theta_3(0,\tau) = \sum_{n\in \mathbb Z} q^{1/2n^2},
\end{eqnarray}
with $q  = e^{2 \pi i \tau}$. 
The action of the $\mathbb{Z}_2$ symmetry sends $u \to - u$, hence we do need a non-trivial fibral transformation $(x,y) \to (x',y')$ in such a way to make $\lambda$ invariant w.r.t. the $\Gamma$-quotient. For a $\mathbb Z_2$ quotient that respects the $U(1)_R$ charges of $x,y$ (hence, that acts diagonally on $x,y$), we have then two-possibilities:
\begin{enumerate}
    \item $x \to -x$ and $y \to y$, but this does not induces an automorphism of the SW geometry \eqref{eq:startingswdatasu2} for any choice of the UV coupling $\tau$;
    \item $x \to x$ and $y \to -y$. 
\end{enumerate}
We see that also the second option is not an automorphism of the SW geometry \eqref{eq:startingswdatasu2} for a generic choice of $\tau$. However, choosing $\tau$ such that, e.g., $e_2 = 0$ (this corresponds to $\tau = i$), we get that
\begin{equation}
    \label{eq:quotientautsu2}
    \widetilde{\Gamma}(x,y,u) = (x,-y,-u)\,,
\end{equation}
preserves both the SW data in \eqref{eq:startingswdatasu2}. We can then use  \eqref{eq:quotientautsu2} to quotient the total SW geometry, and obtain the SW geometry and SW differential for the discrete gauging of the $SU(2) +4 F \oplus \overline{F}$ theory.

Now that we found the candidate $\widetilde{\Gamma}$ action, to perform the quotient we first introduce the invariant coordinates
\begin{equation}
    \label{eq:invcordssu2}
    a = y^2, \quad b \equiv y u, \quad U \equiv u^2
\end{equation}
satisfying the relation
\begin{equation}
    \label{eq:quotienteqsu2}
    a U - b^2 = 0.
\end{equation}
Firstly, we note that the new Coulomb branch coordinate is $U$, hence the projection on $U$ describes the new SW geometry as a fibration of the new Coulomb branch $\mathbb C \ni U$ and is identified with $\widetilde{\pi}$ in \eqref{eq:liftingquotient}. 
Secondly, we rewrite the initial \eqref{eq:startingswdatasu2} SW curve in terms of the invariant coordinates  \eqref{eq:invcordssu2}, obtaining (after a numerical re-scaling of the coordinates),
\begin{equation}
\label{eq:quotientswsu2}
    a +x^3 - U x = 0.
\end{equation}
We can now solve \eqref{eq:quotientswsu2} for $a$ in terms of $(x,U)$, and plug the solution into \eqref{eq:quotienteqsu2}, obtaining
\begin{equation}    \label{eq:finalressu2}
    -b^2 + U x(-U+x^2) = 0.
\end{equation}
We observe that the curve is not in Weierstrass form, but the SW differential is already in the desired \cite{Argyres:2005pp} canonical form $\lambda = u dx/y = U dx/b$. To produce the Weierstrass form, we have to rescale $x \to U^f X, b \to U^g B$. We notice that, since the SW differential is already in the desired canonical form $\lambda = u dx/y = U dx/b$ it must be $f=g$. We then see that the choice $f=-1$ put \eqref{eq:finalressu2} in the Weierstrass form: 
\begin{equation}
    \label{eq:finalresWeierstrasssu2}
    B^2 +X^3 - U^3 X = 0, \qquad \lambda = U \frac{d X}{B}.
\end{equation}
We can observe that \eqref{eq:finalresWeierstrasssu2} is a type $III^*$ Kodaira singularity, as predicted in \cite{Argyres:2016yzz} with slightly different methods.

To conclude and check our result we can count the mass deformations of the theory after the discrete gauging. As in \cite{Argyres:2016yzz}, these are the masses of the parent theory that are invariant w.r.t. the outer-automorphism factor of the discrete gauging. From a SW perspective, we ask ourselves what is the maximum number of mass deformations for which $\widetilde{\Gamma}$ is again an automorphism, or, equivalently, which subvariety of the SW family over the extended Coulomb branch still enjoys $\widetilde{\Gamma}$ as automorphism. For this, we already impose $e_2 = 0$ in the mass-deformed curve of \cite{Seiberg:1994aj}. By explicit computation, one can check that the only term of the mass-deformed SU(2) + 4 F curve that is not $\widetilde{\Gamma}$-invariant is
\begin{equation}
    \label{eq:massdeformedcurvesu2}
    \delta \propto (T_1-T_3)u,
\end{equation}
with
\begin{eqnarray}
\label{eq:masscasimiroriginalpaper}
T_1 = \frac{1}{12}\sum_{i>j}m_i^2m_j^2-\frac{1}{24}\sum_{i} m_i^4, \qquad
T_3 = \frac{1}{2}\left(\prod_{i}m_i - T_1\right), 
\end{eqnarray}
with $i,j = 1,...,4$.
We then see that imposing \eqref{eq:quotientautsu2} to be a symmetry of the mass-deformed geometry forces $T_1  = T_3$ and drops the rank of the flavor group by one. This is compatible with the fact that the new expected flavor algebra is $B_3$ \cite{Argyres:2016yzz}. 
To finely check the group is indeed $B_3$, we follow  \cite{Seiberg:1994aj} and define $T_2 = \frac{1}{2}\left(-\prod_{i}m_i - T_1\right)$. $T_1,T_2,T_3$ are permuted by the triality of the flavor group SO(8), hence one of the three $\mathbb Z_2$ subgroups that folds $D_4$ to $B_3$ exactly exchanges $T_1 \leftrightarrow T_3$. In other words, we see that imposing $T_1 = T_3$ is equivalent to require the masses to lie into a $B_3$ subalgebra of $D_4$, reproducing the expected result \cite{Argyres:2016yzz}.

\section{Explicit check of the monodromy algebra} \label{sec:app_monodromies}
    In this appendix, we will verify that the monodromies that we found in \cref{sec:bpsspectrumsu3tilde} satisfy \eqref{eq:topologicalconditiongeneral}. 
Starting from a base point $u_{6,b}$ with $\text{Re}(u_{6,b}) \to + \infty$, $\text{Im}(u_6,b)=0$, and going anticlockwise we have 

\begin{equation}
    \label{eq:monodromyrelation}
    M_{fixed}M_{p_+}M_{p_{-}}  = M_{loop,u_6}, \qquad M_{loop,u_6} = \left(
\begin{array}{cccc}
 0 & 1 & 0 & 0 \\
 -1 & -1 & 0 & 0 \\
 1 & -1 & -1 & 1 \\
 2 & 1 & -1 & 0 \\
\end{array}
\right)
\end{equation}
One can check that $M_{loop,u_6}^2=M_{loop,u3}$, with 
\begin{equation}
M_{loop,u3}=\left(
\begin{array}{cccc}
 -1 & -1 & 0 & 0 \\
 1 & 0 & 0 & 0 \\
 2 & 4 & 0 & -1 \\
 -2 & 2 & 1 & -1 \\
\end{array}
\right), \qquad M_{loop,u3} = M_{3} M_{4}M_{1}M_{2}
\end{equation}
with $M_j$ being the monodromy around $q_{2,j}$ in \eqref{eq:bpsparticlessu3u2plane}, computed on an anti-clockwise loop starting from a real-valued base-point with large negative real part. We can understand the result by noticing that 
\begin{equation}
M_{loop,u3}=VM_{1}VVM_{2}VM_{1}M_{2} = (VM_{1}M_2)^2 = M_{loop,u6}^2
\end{equation}\\
Now, by using $M_{p_+} = M_{1}, M_{p_-} = M_2$ (that comes from the fact that the point $p_+$ is the $\mathbb Z_2$-quotient image of $u_{2,1}$ etc.), we see that
\begin{equation}
    \label{eq:generalressu3case}
    V M_{1} M_2 = M_{loop,u6} = M_{fixed} M_1 M_2,
\end{equation}
and hence $M_{fixed} = V$. 

\end{appendix}

\printbibliography

@article{Furrer:2024zzu,
    author = "Furrer, Elias and Magureanu, Horia",
    title = "{Coulomb branch surgery: Holonomy saddles, S-folds and discrete symmetry gaugings}",
    eprint = "2404.02955",
    archivePrefix = "arXiv",
    primaryClass = "hep-th",
    month = "4",
    year = "2024"
}

@article{Gaiotto:2014kfa,
    author = "Gaiotto, Davide and Kapustin, Anton and Seiberg, Nathan and Willett, Brian",
    title = "{Generalized Global Symmetries}",
    eprint = "1412.5148",
    archivePrefix = "arXiv",
    primaryClass = "hep-th",
    doi = "10.1007/JHEP02(2015)172",
    journal = "JHEP",
    volume = "02",
    pages = "172",
    year = "2015"
}

@article{Seiberg:2023cdc,
    author = "Seiberg, Nathan and Shao, Shu-Heng",
    title = "{Majorana chain and Ising model -- (non-invertible) translations, anomalies, and emanant symmetries}",
    eprint = "2307.02534",
    archivePrefix = "arXiv",
    primaryClass = "cond-mat.str-el",
    reportNumber = "YITP-SB-2023-14",
    month = "7",
    year = "2023"
}

@article{Karch:2019lnn,
    author = "Karch, Andreas and Tong, David and Turner, Carl",
    title = "{A Web of 2d Dualities: ${\bf Z}_2$ Gauge Fields and Arf Invariants}",
    eprint = "1902.05550",
    archivePrefix = "arXiv",
    primaryClass = "hep-th",
    doi = "10.21468/SciPostPhys.7.1.007",
    journal = "SciPost Phys.",
    volume = "7",
    pages = "007",
    year = "2019"
}

@inproceedings{Martone:2020hvy,
    author = "Martone, Mario",
    title = "{The constraining power of Coulomb Branch Geometry: lectures on Seiberg-Witten theory}",
    booktitle = "{Young Researchers Integrability School and Workshop 2020}: {A modern primer for superconformal field theories}",
    eprint = "2006.14038",
    archivePrefix = "arXiv",
    primaryClass = "hep-th",
    month = "6",
    year = "2020"
}

@article{Argyres:2020wmq,
    author = "Argyres, Philip C. and Martone, Mario",
    title = "{Towards a classification of rank r$ \mathcal{N} $ = 2 SCFTs. Part II. Special Kahler stratification of the Coulomb branch}",
    eprint = "2007.00012",
    archivePrefix = "arXiv",
    primaryClass = "hep-th",
    doi = "10.1007/JHEP12(2020)022",
    journal = "JHEP",
    volume = "12",
    pages = "022",
    year = "2020"
}

@article{Xie:2023lko,
    author = "Xie, Dan",
    title = "{Pseudo-periodic map and classification of theories with eight supercharges}",
    eprint = "2304.13663",
    archivePrefix = "arXiv",
    primaryClass = "hep-th",
    month = "4",
    year = "2023"
}

@article{Caorsi:2018zsq,
    author = "Caorsi, Matteo and Cecotti, Sergio",
    title = "{Geometric classification of 4d $\mathcal{N}=2$ SCFTs}",
    eprint = "1801.04542",
    archivePrefix = "arXiv",
    primaryClass = "hep-th",
    doi = "10.1007/JHEP07(2018)138",
    journal = "JHEP",
    volume = "07",
    pages = "138",
    year = "2018"
}

@article{vanBeest:2022fss,
    author = "van Beest, Marieke and Gould, Dewi S. W. and Schafer-Nameki, Sakura and Wang, Yi-Nan",
    title = "{Symmetry TFTs for 3d QFTs from M-theory}",
    eprint = "2210.03703",
    archivePrefix = "arXiv",
    primaryClass = "hep-th",
    doi = "10.1007/JHEP02(2023)226",
    journal = "JHEP",
    volume = "02",
    pages = "226",
    year = "2023"
}

@article{Apruzzi:2023uma,
    author = "Apruzzi, Fabio and Bonetti, Federico and Gould, Dewi S. W. and Schafer-Nameki, Sakura",
    title = "{Aspects of Categorical Symmetries from Branes: SymTFTs and Generalized Charges}",
    eprint = "2306.16405",
    archivePrefix = "arXiv",
    primaryClass = "hep-th",
    month = "6",
    year = "2023"
}

@article{Bhardwaj:2023ayw,
    author = "Bhardwaj, Lakshya and Schafer-Nameki, Sakura",
    title = "{Generalized Charges, Part II: Non-Invertible Symmetries and the Symmetry TFT}",
    eprint = "2305.17159",
    archivePrefix = "arXiv",
    primaryClass = "hep-th",
    month = "5",
    year = "2023"
}

@article{Chen:2023qnv,
    author = "Chen, Jin and Cui, Wei and Haghighat, Babak and Wang, Yi-Nan",
    title = "{SymTFTs and duality defects from 6d SCFTs on 4-manifolds}",
    eprint = "2305.09734",
    archivePrefix = "arXiv",
    primaryClass = "hep-th",
    doi = "10.1007/JHEP11(2023)208",
    journal = "JHEP",
    volume = "11",
    pages = "208",
    year = "2023"
}

@article{Closset:2023pmc,
    author = "Closset, Cyril and Magureanu, Horia",
    title = "{Reading between the rational sections: Global structures of 4d $\mathcal{N}=2$ KK theories}",
    eprint = "2308.10225",
    archivePrefix = "arXiv",
    primaryClass = "hep-th",
    month = "8",
    year = "2023"
}

@article{Hosseini:2021ged,
    author = "Hosseini, Saghar S. and Moscrop, Robert",
    title = "{Maruyoshi-Song flows and defect groups of $ {\mathrm{D}}_{\mathrm{p}}^{\mathrm{b}} $(G) theories}",
    eprint = "2106.03878",
    archivePrefix = "arXiv",
    primaryClass = "hep-th",
    doi = "10.1007/JHEP10(2021)119",
    journal = "JHEP",
    volume = "10",
    pages = "119",
    year = "2021"
}

@article{Sacchi:2023omn,
    author = "Sacchi, Matteo and Sela, Orr and Zafrir, Gabi",
    title = "{5d to 3d compactifications and discrete anomalies}",
    eprint = "2305.08185",
    archivePrefix = "arXiv",
    primaryClass = "hep-th",
    doi = "10.1007/JHEP10(2023)185",
    journal = "JHEP",
    volume = "10",
    pages = "185",
    year = "2023"
}

@article{Kaidi:2023maf,
    author = "Kaidi, Justin and Nardoni, Emily and Zafrir, Gabi and Zheng, Yunqin",
    title = "{Symmetry TFTs and anomalies of non-invertible symmetries}",
    eprint = "2301.07112",
    archivePrefix = "arXiv",
    primaryClass = "hep-th",
    doi = "10.1007/JHEP10(2023)053",
    journal = "JHEP",
    volume = "10",
    pages = "053",
    year = "2023"
}

@article{Antinucci:2022vyk,
    author = "Antinucci, Andrea and Benini, Francesco and Copetti, Christian and Galati, Giovanni and Rizi, Giovanni",
    title = "{The holography of non-invertible self-duality symmetries}",
    eprint = "2210.09146",
    archivePrefix = "arXiv",
    primaryClass = "hep-th",
    reportNumber = "SISSA 16/2022/FISI",
    month = "10",
    year = "2022"
}

@article{Kaidi:2022cpf,
    author = "Kaidi, Justin and Ohmori, Kantaro and Zheng, Yunqin",
    title = "{Symmetry TFTs for Non-invertible Defects}",
    eprint = "2209.11062",
    archivePrefix = "arXiv",
    primaryClass = "hep-th",
    doi = "10.1007/s00220-023-04859-7",
    journal = "Commun. Math. Phys.",
    volume = "404",
    number = "2",
    pages = "1021--1124",
    year = "2023"
}

@article{Argyres:2022kon,
    author = "Argyres, Philip C. and Martone, Mario and Ray, Michael",
    title = "{Dirac pairings, one-form symmetries and Seiberg-Witten geometries}",
    eprint = "2204.09682",
    archivePrefix = "arXiv",
    primaryClass = "hep-th",
    doi = "10.1007/JHEP09(2022)020",
    journal = "JHEP",
    volume = "09",
    pages = "020",
    year = "2022"
}

@article{DelZotto:2022ras,
    author = "Del Zotto, Michele and Garc\'\i{}a Etxebarria, I\~naki",
    title = "{Global structures from the infrared}",
    eprint = "2204.06495",
    archivePrefix = "arXiv",
    primaryClass = "hep-th",
    doi = "10.1007/JHEP11(2023)058",
    journal = "JHEP",
    volume = "11",
    pages = "058",
    year = "2023"
}

@article{Apruzzi:2022dlm,
    author = "Apruzzi, Fabio",
    title = "{Higher form symmetries TFT in 6d}",
    eprint = "2203.10063",
    archivePrefix = "arXiv",
    primaryClass = "hep-th",
    doi = "10.1007/JHEP11(2022)050",
    journal = "JHEP",
    volume = "11",
    pages = "050",
    year = "2022"
}

@article{DelZotto:2022joo,
    author = "Del Zotto, Michele and Garc\'\i{}a Etxebarria, I\~naki and Schafer-Nameki, Sakura",
    title = "{2-Group Symmetries and M-Theory}",
    eprint = "2203.10097",
    archivePrefix = "arXiv",
    primaryClass = "hep-th",
    doi = "10.21468/SciPostPhys.13.5.105",
    journal = "SciPost Phys.",
    volume = "13",
    pages = "105",
    year = "2022"
}

@article{Apruzzi:2021nmk,
    author = "Apruzzi, Fabio and Bonetti, Federico and Garc\'\i{}a Etxebarria, I\~naki and Hosseini, Saghar S. and Schafer-Nameki, Sakura",
    title = "{Symmetry TFTs from String Theory}",
    eprint = "2112.02092",
    archivePrefix = "arXiv",
    primaryClass = "hep-th",
    doi = "10.1007/s00220-023-04737-2",
    journal = "Commun. Math. Phys.",
    volume = "402",
    number = "1",
    pages = "895--949",
    year = "2023"
}

@article{Xie:2023wqx,
    author = "Xie, Dan",
    title = "{On rank two theories with eight supercharges part II: Lefschetz pencils}",
    eprint = "2311.15986",
    archivePrefix = "arXiv",
    primaryClass = "hep-th",
    month = "11",
    year = "2023"
}

@article{Li:2022njl,
    author = "Li, Bohan and Xie, Dan and Yan, Wenbin",
    title = "{On low rank 4d $ \mathcal{N} $ = 2 SCFTs}",
    eprint = "2212.03089",
    archivePrefix = "arXiv",
    primaryClass = "hep-th",
    doi = "10.1007/JHEP05(2023)132",
    journal = "JHEP",
    volume = "05",
    pages = "132",
    year = "2023"
}

@article{Xie:2022aad,
    author = "Xie, Dan",
    title = "{On rank two theories with eight supercharges part I: local singularities}",
    eprint = "2212.02472",
    archivePrefix = "arXiv",
    primaryClass = "hep-th",
    month = "12",
    year = "2022"
}

@article{Argyres:2022lah,
    author = "Argyres, Philip C. and Martone, Mario",
    title = "{The rank 2 classification problem I: scale invariant geometries}",
    eprint = "2209.09248",
    archivePrefix = "arXiv",
    primaryClass = "hep-th",
    month = "9",
    year = "2022"
}

@article{Argyres:2022fwy,
    author = "Argyres, Philip C. and Martone, Mario",
    title = "{The rank-2 classification problem III: curves with additional automorphisms}",
    eprint = "2209.10555",
    archivePrefix = "arXiv",
    primaryClass = "hep-th",
    month = "9",
    year = "2022"
}

@article{Argyres:2022puv,
    author = "Argyres, Philip C. and Martone, Mario",
    title = "{The rank 2 classification problem II: mapping scale-invariant solutions to SCFTs}",
    eprint = "2209.09911",
    archivePrefix = "arXiv",
    primaryClass = "hep-th",
    month = "9",
    year = "2022"
}

@article{Argyres:2016xmc,
    author = {Argyres, Philp and Lotito, Matteo and L\"u, Yongchao and Martone, Mario},
    title = "{Geometric constraints on the space of $ \mathcal{N}$ = 2 SCFTs. Part III: enhanced Coulomb branches and central charges}",
    eprint = "1609.04404",
    archivePrefix = "arXiv",
    primaryClass = "hep-th",
    doi = "10.1007/JHEP02(2018)003",
    journal = "JHEP",
    volume = "02",
    pages = "003",
    year = "2018"
}

@article{Argyres:2016xua,
    author = {Argyres, Philip C. and Lotito, Matteo and L\"u, Yongchao and Martone, Mario},
    title = "{Expanding the landscape of $ \mathcal{N} $ = 2 rank 1 SCFTs}",
    eprint = "1602.02764",
    archivePrefix = "arXiv",
    primaryClass = "hep-th",
    doi = "10.1007/JHEP05(2016)088",
    journal = "JHEP",
    volume = "05",
    pages = "088",
    year = "2016"
}

@article{Argyres:2015gha,
    author = {Argyres, Philip C. and Lotito, Matteo and L\"u, Yongchao and Martone, Mario},
    title = {{Geometric constraints on the space of $ \mathcal{N} $ = 2 SCFTs. Part II: construction of special K\"ahler geometries and RG flows}},
    eprint = "1601.00011",
    archivePrefix = "arXiv",
    primaryClass = "hep-th",
    doi = "10.1007/JHEP02(2018)002",
    journal = "JHEP",
    volume = "02",
    pages = "002",
    year = "2018"
}

@article{Argyres:2015ffa,
    author = {Argyres, Philip and Lotito, Matteo and L\"u, Yongchao and Martone, Mario},
    title = "{Geometric constraints on the space of $ \mathcal{N} $ = 2 SCFTs. Part I: physical constraints on relevant deformations}",
    eprint = "1505.04814",
    archivePrefix = "arXiv",
    primaryClass = "hep-th",
    doi = "10.1007/JHEP02(2018)001",
    journal = "JHEP",
    volume = "02",
    pages = "001",
    year = "2018"
}

@article{Harlow:2023hjb,
    author = "Harlow, Daniel and Numasawa, Tokiro",
    title = "{Gauging spacetime inversions in quantum gravity}",
    eprint = "2311.09978",
    archivePrefix = "arXiv",
    primaryClass = "hep-th",
    reportNumber = "MIT-CTP/5647",
    month = "11",
    year = "2023"
}

@article{Harlow:2018tng,
    author = "Harlow, Daniel and Ooguri, Hirosi",
    title = "{Symmetries in quantum field theory and quantum gravity}",
    eprint = "1810.05338",
    archivePrefix = "arXiv",
    primaryClass = "hep-th",
    doi = "10.1007/s00220-021-04040-y",
    journal = "Commun. Math. Phys.",
    volume = "383",
    number = "3",
    pages = "1669--1804",
    year = "2021"
}

@article{Benini:2018reh,
    author = "Benini, Francesco and C\'ordova, Clay and Hsin, Po-Shen",
    title = "{On 2-Group Global Symmetries and their Anomalies}",
    eprint = "1803.09336",
    archivePrefix = "arXiv",
    primaryClass = "hep-th",
    reportNumber = "SISSA 10/2018/FISI, SISSA-10-2018-FISI",
    doi = "10.1007/JHEP03(2019)118",
    journal = "JHEP",
    volume = "03",
    pages = "118",
    year = "2019"
}

@article{Bhardwaj:2022scy,
    author = "Bhardwaj, Lakshya and Gould, Dewi S. W.",
    title = "{Disconnected 0-form and 2-group symmetries}",
    eprint = "2206.01287",
    archivePrefix = "arXiv",
    primaryClass = "hep-th",
    doi = "10.1007/JHEP07(2023)098",
    journal = "JHEP",
    volume = "07",
    pages = "098",
    year = "2023"
}

@article{Kallosh:1995hi,
    author = "Kallosh, Renata and Linde, Andrei D. and Linde, Dmitri A. and Susskind, Leonard",
    title = "{Gravity and global symmetries}",
    eprint = "hep-th/9502069",
    archivePrefix = "arXiv",
    reportNumber = "SU-ITP-95-2",
    doi = "10.1103/PhysRevD.52.912",
    journal = "Phys. Rev. D",
    volume = "52",
    pages = "912--935",
    year = "1995"
}

@article{Arias-Tamargo:2019jyh,
    author = "Arias-Tamargo, Guillermo and Bourget, Antoine and Pini, Alessandro and Rodr\'\i{}guez-G\'omez, Diego",
    title = "{Discrete gauge theories of charge conjugation}",
    eprint = "1903.06662",
    archivePrefix = "arXiv",
    primaryClass = "hep-th",
    reportNumber = "DESY-19-045, Imperial/TP/2019/AB/01",
    doi = "10.1016/j.nuclphysb.2019.114721",
    journal = "Nucl. Phys. B",
    volume = "946",
    pages = "114721",
    year = "2019"
}

@article{Arias-Tamargo:2021ppf,
    author = "Arias-Tamargo, Guillermo and Bourget, Antoine and Pini, Alessandro",
    title = "{Discrete gauging and Hasse diagrams}",
    eprint = "2105.08755",
    archivePrefix = "arXiv",
    primaryClass = "hep-th",
    doi = "10.21468/SciPostPhys.11.2.026",
    journal = "SciPost Phys.",
    volume = "11",
    number = "2",
    pages = "026",
    year = "2021"
}

@book{brocker2013representations,
  title={Representations of compact Lie groups},
  author={Br{\"o}cker, Theodor and Tom Dieck, Tammo},
  volume={98},
  year={2013},
  publisher={Springer Science \& Business Media}
}

@article{Aharony:2013hda,
    author = "Aharony, Ofer and Seiberg, Nathan and Tachikawa, Yuji",
    title = "{Reading between the lines of four-dimensional gauge theories}",
    eprint = "1305.0318",
    archivePrefix = "arXiv",
    primaryClass = "hep-th",
    reportNumber = "WIS-03-13-APR-DPPA, WIS/03/13-APR-DPPA, UT-13-15, IPMU13-0081",
    doi = "10.1007/JHEP08(2013)115",
    journal = "JHEP",
    volume = "08",
    pages = "115",
    year = "2013"
}

@article{Hanany:1995na,
    author = "Hanany, Amihay and Oz, Yaron",
    title = "{On the quantum moduli space of vacua of N=2 supersymmetric SU(N(c)) gauge theories}",
    eprint = "hep-th/9505075",
    archivePrefix = "arXiv",
    reportNumber = "TAUP-2248-95, WIS-95-19-PH",
    doi = "10.1016/0550-3213(95)00376-4",
    journal = "Nucl. Phys. B",
    volume = "452",
    pages = "283--312",
    year = "1995"
}

@book{arnold1985singularities2,
  title={Singularities of Differentiable Maps: Volume II},
  author={Arnold, V.I. and Varchenko, A. and Gusein-Zade, S.M.},
  isbn={9780817631871},
  lccn={lc84012134},
  series={Monographs in Mathematics},
  url={https://books.google.it/books?id=kceNLJpmMiUC},
  year={1985},
  publisher={Birkh{\"a}user Boston}
}

@article{Lerche_1997,
   title={Introduction to seiberg-witten theory and its stringy origin},
   volume={55},
   ISSN={0920-5632},
   url={http://dx.doi.org/10.1016/S0920-5632(97)00073-X},
   DOI={10.1016/s0920-5632(97)00073-x},
   number={2},
   journal={Nuclear Physics B - Proceedings Supplements},
   publisher={Elsevier BV},
   author={Lerche, W.},
   year={1997},
   month={5},
   pages={83–117}
}

@article{Argyres:2018wxu,
    author = "Argyres, Philip C. and Martone, Mario",
    title = "{Coulomb branches with complex singularities}",
    eprint = "1804.03152",
    archivePrefix = "arXiv",
    primaryClass = "hep-th",
    doi = "10.1007/JHEP06(2018)045",
    journal = "JHEP",
    volume = "06",
    pages = "045",
    year = "2018"
}

@article{Banks_2011,
   title={Symmetries and strings in field theory and gravity},
   volume=83,
   ISSN={1550-2368},
   url={http://dx.doi.org/10.1103/PhysRevD.83.084019},
   DOI={10.1103/physrevd.83.084019},
   number=8,
   journal={Physical Review D},
   publisher={American Physical Society (APS)},
   author={Banks, Tom and Seiberg, Nathan},
   year=2011,
   month={4}
}

@article{Argyres:1995wt,
    author = "Argyres, Philip C. and Plesser, M. Ronen and Shapere, Alfred D.",
    title = "{The Coulomb phase of N=2 supersymmetric QCD}",
    eprint = "hep-th/9505100",
    archivePrefix = "arXiv",
    reportNumber = "IASSNS-HEP-95-32, UK-HEP-95-06",
    doi = "10.1103/PhysRevLett.75.1699",
    journal = "Phys. Rev. Lett.",
    volume = "75",
    pages = "1699--1702",
    year = "1995"
}

@article{Argyres:2005pp,
    author = "Argyres, Philip C. and Crescimanno, Michael and Shapere, Alfred D. and Wittig, John R.",
    title = "{Classification of N=2 superconformal field theories with two-dimensional Coulomb branches}",
    eprint = "hep-th/0504070",
    archivePrefix = "arXiv",
    month = "4",
    year = "2005"
}

@article{Seiberg:1994aj,
    author = "Seiberg, N. and Witten, Edward",
    title = "{Monopoles, duality and chiral symmetry breaking in N=2 supersymmetric QCD}",
    eprint = "hep-th/9408099",
    archivePrefix = "arXiv",
    reportNumber = "RU-94-60, IASSNS-HEP-94-55",
    doi = "10.1016/0550-3213(94)90214-3",
    journal = "Nucl. Phys. B",
    volume = "431",
    pages = "484--550",
    year = "1994"
}

@article{Bhardwaj:2021zrt,
    author = "Bhardwaj, Lakshya and Hubner, Max and Schafer-Nameki, Sakura",
    title = "{Liberating confinement from Lagrangians: 1-form symmetries and lines in 4d N=1 from 6d N=(2,0)}",
    eprint = "2106.10265",
    archivePrefix = "arXiv",
    primaryClass = "hep-th",
    doi = "10.21468/SciPostPhys.12.1.040",
    journal = "SciPost Phys.",
    volume = "12",
    number = "1",
    pages = "040",
    year = "2022"
}

@article{Alastair:2002yt,
    author = "Alastair, Craw and Akira, Ishii",
    title = "{Flops of G Hilb and equivalences of derived categories by variation of GIT quotient}",
    eprint = "math/0211360",
    archivePrefix = "arXiv",
    doi = "10.1215/S0012-7094-04-12422-4",
    journal = "Duke Math. J.",
    volume = "124",
    pages = "259--307",
    year = "2004"
}

@article{Argyres:2016yzz,
    author = "Argyres, Philip C. and Martone, Mario",
    title = "{4d $ \mathcal{N} $ =2 theories with disconnected gauge groups}",
    eprint = "1611.08602",
    archivePrefix = "arXiv",
    primaryClass = "hep-th",
    doi = "10.1007/JHEP03(2017)145",
    journal = "JHEP",
    volume = "03",
    pages = "145",
    year = "2017"
}

@article{Collinucci:2022rii,
    author = "Collinucci, Andr\'es and De Marco, Mario and Sangiovanni, Andrea and Valandro, Roberto",
    title = "{Flops of any length, Gopakumar-Vafa invariants and 5d Higgs branches}",
    eprint = "2204.10366",
    archivePrefix = "arXiv",
    primaryClass = "hep-th",
    doi = "10.1007/JHEP08(2022)292",
    journal = "JHEP",
    volume = 08,
    pages = 292,
    year = 2022
}

@article{DeMarco:2022dgh,
    author = "De Marco, Mario and Sangiovanni, Andrea and Valandro, Roberto",
    title = "{5d Higgs branches from M-theory on quasi-homogeneous cDV threefold singularities}",
    eprint = "2205.01125",
    archivePrefix = "arXiv",
    primaryClass = "hep-th",
    doi = "10.1007/JHEP10(2022)124",
    journal = "JHEP",
    volume = "10",
    pages = "124",
    year = "2022"
}

@article{Martone:2020nsy,
    author = "Martone, Mario",
    title = "{Towards the classification of rank-r$ \mathcal{N} $ = 2 SCFTs. Part I. Twisted partition function and central charge formulae}",
    eprint = "2006.16255",
    archivePrefix = "arXiv",
    primaryClass = "hep-th",
    doi = "10.1007/JHEP12(2020)021",
    journal = "JHEP",
    volume = "12",
    pages = "021",
    year = "2020"
}

@article{DeMarco:2021try,
    author = "De Marco, Mario and Sangiovanni, Andrea",
    title = "{Higgs Branches of rank-0 5d theories from M-theory on (A$_{j}$, A$_{l}$) and (A$_{k}$, D$_{n}$) singularities}",
    eprint = "2111.05875",
    archivePrefix = "arXiv",
    primaryClass = "hep-th",
    doi = "10.1007/JHEP03(2022)099",
    journal = "JHEP",
    volume = "03",
    pages = "099",
    year = "2022"
}

@article{Collinucci:2021ofd,
    author = "Collinucci, Andr\'es and De Marco, Mario and Sangiovanni, Andrea and Valandro, Roberto",
    title = "{Higgs branches of 5d rank-zero theories from geometry}",
    eprint = "2105.12177",
    archivePrefix = "arXiv",
    primaryClass = "hep-th",
    doi = "10.1007/JHEP10(2021)018",
    journal = "JHEP",
    volume = "10",
    number = "18",
    pages = "018",
    year = "2021"
}

@inproceedings{Klemm:1994qj,
    author = "Klemm, A. and Lerche, W. and Yankielowicz, S. and Theisen, S.",
    title = "{On the monodromies of N=2 supersymmetric Yang-Mills theory}",
    booktitle = "{Joint U.S.-Polish Workshop on Physics from Planck Scale to Electro-Weak Scale (SUSY 94)}",
    eprint = "hep-th/9412158",
    archivePrefix = "arXiv",
    reportNumber = "CERN-TH-7538-94, LMU-TPW-94-22",
    month = "8",
    year = "1994"
}

@article{Klemm:1995wp,
    author = "Klemm, A. and Lerche, W. and Theisen, S.",
    title = "{Nonperturbative effective actions of N=2 supersymmetric gauge theories}",
    eprint = "hep-th/9505150",
    archivePrefix = "arXiv",
    reportNumber = "CERN-TH-95-104, LMU-TPW-95-7",
    doi = "10.1142/S0217751X96001000",
    journal = "Int. J. Mod. Phys. A",
    volume = "11",
    pages = "1929--1974",
    year = "1996"
}

@article{Ceresole:1993nz,
    author = "Ceresole, Anna and D'Auria, R. and Regge, T.",
    title = "{Duality group for Calabi-Yau 2 moduli space}",
    eprint = "hep-th/9307151",
    archivePrefix = "arXiv",
    reportNumber = "DFTT-34-93, POLFIS-TH-05-93",
    doi = "10.1016/0550-3213(94)90439-1",
    journal = "Nucl. Phys. B",
    volume = "414",
    pages = "517--537",
    year = "1994"
}

@article{Gaiotto:2009we,
    author = "Gaiotto, Davide",
    title = "{N=2 dualities}",
    eprint = "0904.2715",
    archivePrefix = "arXiv",
    primaryClass = "hep-th",
    doi = "10.1007/JHEP08(2012)034",
    journal = "JHEP",
    volume = "08",
    pages = "034",
    year = "2012"
}

@article{GarciaEtxebarria:2022vzq,
    author = "Garc\'\i{}a Etxebarria, I\~naki",
    title = "{Branes and Non-Invertible Symmetries}",
    eprint = "2208.07508",
    archivePrefix = "arXiv",
    primaryClass = "hep-th",
    doi = "10.1002/prop.202200154",
    journal = "Fortsch. Phys.",
    volume = "70",
    number = "11",
    pages = "2200154",
    year = "2022"
}

@article{Damia:2023gtc,
    author = "Damia, Jeremias Aguilera and Argurio, Riccardo and Chaudhuri, Soumyadeep",
    title = "{When the moduli space is an orbifold: Spontaneous breaking of continuous non-invertible symmetries}",
    eprint = "2309.06491",
    archivePrefix = "arXiv",
    primaryClass = "hep-th",
    month = "9",
    year = "2023"
}

@article{Bergman:2022otk,
    author = "Bergman, Oren and Hirano, Shinji",
    title = "{The holography of duality in $ \mathcal{N} $ = 4 Super-Yang-Mills theory}",
    eprint = "2208.09396",
    archivePrefix = "arXiv",
    primaryClass = "hep-th",
    doi = "10.1007/JHEP11(2022)069",
    journal = "JHEP",
    volume = "11",
    pages = "069",
    year = "2022"
}

@article{Heidenreich:2021xpr,
    author = "Heidenreich, Ben and McNamara, Jacob and Montero, Miguel and Reece, Matthew and Rudelius, Tom and Valenzuela, Irene",
    title = "{Non-invertible global symmetries and completeness of the spectrum}",
    eprint = "2104.07036",
    archivePrefix = "arXiv",
    primaryClass = "hep-th",
    reportNumber = "ACFI-T21-03",
    doi = "10.1007/JHEP09(2021)203",
    journal = "JHEP",
    volume = "09",
    pages = "203",
    year = "2021"
}

@article{Tachikawa:2017gyf,
    author = "Tachikawa, Yuji",
    title = "{On gauging finite subgroups}",
    eprint = "1712.09542",
    archivePrefix = "arXiv",
    primaryClass = "hep-th",
    reportNumber = "IPMU-17-0183",
    doi = "10.21468/SciPostPhys.8.1.015",
    journal = "SciPost Phys.",
    volume = "8",
    number = "1",
    pages = "015",
    year = "2020"
}

@article{Katz:1997eq,
    author = "Katz, S. and Mayr, P. and Vafa, C.",
    title = "{Mirror symmetry and exact solution of 4-D N=2 gauge theories: 1.}",
    eprint = "hep-th/9706110",
    archivePrefix = "arXiv",
    reportNumber = "HUTP-97-A025, OSU-M-97-5, IASSNS-HEP-97-65",
    doi = "10.4310/ATMP.1997.v1.n1.a2",
    journal = "Adv. Theor. Math. Phys.",
    volume = "1",
    pages = "53--114",
    year = "1998"
}

@article{Tachikawa:2011yr,
    author = "Tachikawa, Yuji and Terashima, Seiji",
    title = "{Seiberg-Witten Geometries Revisited}",
    eprint = "1108.2315",
    archivePrefix = "arXiv",
    primaryClass = "hep-th",
    reportNumber = "YITP-11-73, IPMU11-0130",
    doi = "10.1007/JHEP09(2011)010",
    journal = "JHEP",
    volume = "09",
    pages = "010",
    year = "2011"
}

@article{Giacomelli:2024sex,
    author = "Giacomelli, Simone and Harding, William and Mekareeya, Noppadol and Mininno, Alessandro",
    title = "{Discrete Global Symmetries: Gauging and Twisted Compactification}",
    eprint = "2402.03424",
    archivePrefix = "arXiv",
    primaryClass = "hep-th",
    reportNumber = "ZMP-HH/24-2",
    month = "2",
    year = "2024"
}

@book{Tachikawa:2013kta,
    author = "Tachikawa, Yuji",
    title = "{N=2 supersymmetric dynamics for pedestrians}",
    eprint = "1312.2684",
    archivePrefix = "arXiv",
    primaryClass = "hep-th",
    reportNumber = "UT-13-42, IPMU-13-0234",
    doi = "10.1007/978-3-319-08822-8",
    month = "12",
    year = "2013"
}

@article{Witten:1997sc,
    author = "Witten, Edward",
    title = "{Solutions of four-dimensional field theories via M theory}",
    eprint = "hep-th/9703166",
    archivePrefix = "arXiv",
    reportNumber = "IASSNS-HEP-97-19",
    doi = "10.1016/S0550-3213(97)00416-1",
    journal = "Nucl. Phys. B",
    volume = "500",
    pages = "3--42",
    year = "1997"
}

@article{Akhond:2021xio,
    author = "Akhond, Mohammad and Arias-Tamargo, Guillermo and Mininno, Alessandro and Sun, Hao-Yu and Sun, Zhengdi and Wang, Yifan and Xu, Fengjun",
    title = "{The hitchhiker's guide to 4d $\mathcal{N}=2$ superconformal field theories}",
    eprint = "2112.14764",
    archivePrefix = "arXiv",
    primaryClass = "hep-th",
    reportNumber = "IFT-UAM/CSIC-21-151, ZMP-HH/21-28",
    doi = "10.21468/SciPostPhysLectNotes.64",
    journal = "SciPost Phys. Lect. Notes",
    volume = "64",
    pages = "1",
    year = "2022"
}

@article{Arias-Tamargo:2022nlf,
    author = "Arias-Tamargo, Guillermo and Rodriguez-Gomez, Diego",
    title = "{Non-invertible symmetries from discrete gauging and completeness of the spectrum}",
    eprint = "2204.07523",
    archivePrefix = "arXiv",
    primaryClass = "hep-th",
    doi = "10.1007/JHEP04(2023)093",
    journal = "JHEP",
    volume = "04",
    pages = "093",
    year = "2023"
}

@article{Seiberg:1994rs,
    author = "Seiberg, N. and Witten, Edward",
    title = "{Electric - magnetic duality, monopole condensation, and confinement in N=2 supersymmetric Yang-Mills theory}",
    eprint = "hep-th/9407087",
    archivePrefix = "arXiv",
    reportNumber = "RU-94-52, IASSNS-HEP-94-43",
    doi = "10.1016/0550-3213(94)90124-4",
    journal = "Nucl. Phys. B",
    volume = "426",
    pages = "19--52",
    year = "1994",
    note = "[Erratum: Nucl.Phys.B 430, 485--486 (1994)]"
}

@article{Shao:2023gho,
    author = "Shao, Shu-Heng",
    title = "{What's Done Cannot Be Undone: TASI Lectures on Non-Invertible Symmetry}",
    eprint = "2308.00747",
    archivePrefix = "arXiv",
    primaryClass = "hep-th",
    reportNumber = "YITP-SB-2023-19",
    month = "8",
    year = "2023"
}

@inproceedings{Carqueville:2023jhb,
    author = "Carqueville, Nils and Del Zotto, Michele and Runkel, Ingo",
    title = "{Topological defects}",
    eprint = "2311.02449",
    archivePrefix = "arXiv",
    primaryClass = "math-ph",
    month = "11",
    year = "2023"
}

@article{Bourget:2018ond,
    author = "Bourget, Antoine and Pini, Alessandro and Rodr\'\i{}guez-G\'omez, Diego",
    title = "{Gauge theories from principally extended disconnected gauge groups}",
    eprint = "1804.01108",
    archivePrefix = "arXiv",
    primaryClass = "hep-th",
    reportNumber = "DESY-18-046",
    doi = "10.1016/j.nuclphysb.2019.02.004",
    journal = "Nucl. Phys. B",
    volume = "940",
    pages = "351--376",
    year = "2019"
}

@article{Brennan:2023mmt,
    author = "Brennan, T. Daniel and Hong, Sungwoo",
    title = "{Introduction to Generalized Global Symmetries in QFT and Particle Physics}",
    eprint = "2306.00912",
    archivePrefix = "arXiv",
    primaryClass = "hep-ph",
    month = "6",
    year = "2023"
}

@article{Bhardwaj:2023kri,
    author = "Bhardwaj, Lakshya and Bottini, Lea E. and Fraser-Taliente, Ludovic and Gladden, Liam and Gould, Dewi S. W. and Platschorre, Arthur and Tillim, Hannah",
    title = "{Lectures on Generalized Symmetries}",
    eprint = "2307.07547",
    archivePrefix = "arXiv",
    primaryClass = "hep-th",
    month = "7",
    year = "2023"
}

@article{Schafer-Nameki:2023jdn,
    author = "Schafer-Nameki, Sakura",
    title = "{ICTP Lectures on (Non-)Invertible Generalized Symmetries}",
    eprint = "2305.18296",
    archivePrefix = "arXiv",
    primaryClass = "hep-th",
    month = "5",
    year = "2023"
}

@article{Antinucci:2022eat,
    author = "Antinucci, Andrea and Galati, Giovanni and Rizi, Giovanni",
    title = "{On continuous 2-category symmetries and Yang-Mills theory}",
    eprint = "2206.05646",
    archivePrefix = "arXiv",
    primaryClass = "hep-th",
    doi = "10.1007/JHEP12(2022)061",
    journal = "JHEP",
    volume = "12",
    pages = "061",
    year = "2022"
}

@article{Bhardwaj:2022yxj,
    author = "Bhardwaj, Lakshya and Bottini, Lea E. and Schafer-Nameki, Sakura and Tiwari, Apoorv",
    title = "{Non-invertible higher-categorical symmetries}",
    eprint = "2204.06564",
    archivePrefix = "arXiv",
    primaryClass = "hep-th",
    doi = "10.21468/SciPostPhys.14.1.007",
    journal = "SciPost Phys.",
    volume = "14",
    number = "1",
    pages = "007",
    year = "2023"
}

\end{document}